\documentclass[%
 reprint,
 amsmath,amssymb,
 aps,
floatfix,
]{revtex4-2}

\usepackage{xcolor}

\usepackage{url}
\usepackage[hyperindex,breaklinks]{hyperref}
\hypersetup{linkcolor=blue, citecolor=red, colorlinks=true}

\usepackage{braket}
\usepackage{physics}
\usepackage{mathtools}
\usepackage{amsfonts}

\usepackage{graphicx}
\usepackage{dcolumn}
\usepackage{bm}
\usepackage{diagbox}

\usepackage{float}

\usepackage{placeins} 

\usepackage{soul}

\begin{document}
\preprint{APS/123-QED}


\title{Enhancing entanglement with the generalized elephant quantum walk from localized and delocalized states}



\author{Caio B. Naves$^{1}$}
\thanks{caio.naves@usp.br}

\author{Marcelo A. Pires$^{2}$}
\thanks{piresma@cbpf.br}

\author{Diogo O. Soares-Pinto$^{1}$}
\thanks{dosp@ifsc.usp.br}

\author{S\'{\i}lvio M. \surname{Duarte~Queir\'{o}s}$^{3}$}
\thanks{sdqueiro@cbpf.br, On leave of absence from CBPF, Brazil} 

\affiliation{
$^{1}$Instituto de F\'{\i}sica de S\~ao Carlos, Universidade de S\~ao Paulo, CP 369, 13560-970, S\~ao Carlos, SP, Brazil
\\
$^{2}$Departamento de Física, Universidade Federal do Cear\'a, 60451-970,  Fortaleza, Brazil
\\
$^{3}$National Institute of Science and Technology for Complex Systems, Brazil
}

\date{\today}

\begin{abstract}

Recently, it was introduced a generalization  of a nonstandard step operator named the elephant quantum walk (EQW). With proper statistical distribution for the steps, that generalized EQW (gEQW) can be tuned to exhibit a myriad of dynamical scaling behavior ranging from standard diffusion to 
hyperballistic spreading. 
In this work, we study the influence of the statistics of the step size and the delocalization of the initial states on the entanglement entropy of the coin. Our results show that the gEQW  generates maximally entangled states for almost all initial coin states and coin operators considering initially localized walkers and for the delocalized ones, taking the proper limit, the same condition is guaranteed.
 Differently from all the previous protocols that produce  highly  entangled  states via QWs, this model is not upper-bounded  by ballistic spreading and hence opens novel prospects for applications of dynamically disordered QWs as a robust maximal entanglement generator in programmable setups that ranges from  slower-than-ballistic to faster-than ballistic. 
\end{abstract} 



\maketitle


\section{\label{sec:intro}Introduction}

By making use of the quantum principle of superposition, the space-time evolution of a quantum particle -- a walker -- on a grid is understood to be as exponentially more powerful than its inspiring classical random walk analogue~\cite{aharonov1993quantum,PhysicsWorld2009, widera2015walk} in scientific and technological applications such as search algorithms and transport phenomena, namely photosynthetic energy\cite{Lloyd2008,venegas2008quantum,portugal2013quantum}, coherent transport~\cite{ambainis2003quantum,kempe2003quantum,kendon2006random,kadian2021quantum,kitagawa2012topological,wu2019topological} among many other implementations either experimental~\cite{grafe2016integrated,grafe2020integrated,wang2013physical,neves2018photonic} or closely related to applications of quantum computation~\cite{optionpricing2020}. 

Since the introduction of the first quantum walk (QW)~\cite{aharonov1993quantum}, a myriad of variants and extensions have been introduced (see references above); primarily, they aimed at studying diffusion-scale properties and search performance enhancing regarding the optimal Grover's algorithm~\cite{Grover1996}. In being a quantum system described by the composition of two sub-spaces  -- i.e., walker and coin --  , the QW problem is naturally suitable to the analysis of a pure quantum feature: entanglement. Nonetheless, the measurement between its sub-spaces only took place some years afterwards~\cite{carneiro2005entanglement}.
On the other hand, quantum features do not boil down to entanglement; e.g., the localization conditions is another crucial property for the understanding of quantum effects~\cite{brandes2003}.

In the present manuscript, we analyze the outcome on the entanglement entropy for a general disordered quantum walk model that is able to describe a wide range of dynamical features -- from normal (like a classical random walker) to hyperballistic diffusion, as experimentally verified in hybrid ordered–disordered photonic lattices~\cite{stutzer2013superballistic} -- considering different localization status.

The remaining of this text is organized as follows: in Sec.~\ref{sec:lit-rev} we provide a concise review over the subject-matters we have aforementioned and which are at the core of the present study; in Sec.~\ref{sec:model} we introduce the model and in Sec.~\ref{sec:results} we report our results and discuss them. At last, in Sec.~\ref{sec:remarks} we address some concluding words in our study and point out further avenues of research.

\section{\label{sec:lit-rev}Literature review}

The canonical discrete-time quantum walk was pioneered in 1993 \cite{aharonov1993quantum}. It is a lattice-based model exhibiting interesting features~\cite{kempe2003quantum,kendon2006random,kadian2021quantum,ambainis2003quantum,venegas2008quantum,portugal2013quantum} when compared with the classical random walk (CRW). First, in its standard version, the QW is quadratically faster than the CRW, second, it shows a non-Gaussian bimodal distribution while the CRW displays the usual Gaussian distribution. Apart from this, the QW is able to entangle their internal (spin) and external (position) degrees of freedom. This  feature can be quantified by the entanglement entropy $S_E$ whose value was firstly numerically estimated in 0.872 in the asymptotic regime~\cite{carneiro2005entanglement}. Such a value was posteriorly demonstrated mathematically~\cite{abal2006quantum} and experimentally~\cite{wang2018dynamic,tao2021experimental,su2019experimental}.

The nonstandard versions of the QW are even more interesting. For instance, in 2012~\cite{chandrashekar2012disorder} it was numerically shown that QWs with random temporal disorder can be used as a maximal entanglement generator. This counterintuitive result was  demonstrated mathematically in 2013~\cite{vieira2013dynamically} and experimentally in 2018~\cite{wang2018dynamic}.  
Further works have provided an analysis of the entanglement considering several protocols of disorder embedded either in the coin operator~\cite{salimi2012asymptotic,rohde2013quantum,vieira2014entangling,di2016discrete,orthey2019weak,singh2019accelerated,montero2016classical,zeng2017discrete,buarque2019aperiodic,pires2020genuine,gratsea2020generation,gratsea2020universal,walczak2021parrondo,pires2021negative,zhang2022maximal} or in the step operator~\cite{pires2019multiple,sen2020scaling,pires2020quantum}.  With some exceptions, the overwhelming majority of these works focused on QW dynamics from local initial states. In Ref.\cite{orthey2017asymptotic} it was shown that for disorder-free QWs delocalization increases the number of initial states which reach maximal asymptotic entanglement.

At this point a question arises: How does the entanglement in QWs change in the presence of both disorder and delocalization? This issue was briefly and partially addressed when temporal disorder was embedded in the coin operator~\cite{vieira2014entangling} where it was shown that the maximal asymptotic entanglement is still achieved. However, the aforementioned question is still open when temporal disorder is introduced in the step operator.

Even though QWs with temporal disorder in the coin operator
have been studied for a much longer time~\cite{kempe2003quantum,kendon2006random,kadian2021quantum,ambainis2003quantum,venegas2008quantum,portugal2013quantum}, the 
QWs with temporal disorder in the displacement operator exhibits an interesting phenomenology~\cite{mulken2008universal,caceres2010quantum,lavivcka2011quantum,zhao2015one,chattaraj2016effects,di2018elephant,pires2019multiple,das2019inhibition,sen2019unusual,sen2020scaling,mukhopadhyay2020persistent,zaman2022randomizing,pires2020quantum,dominguez2021enhanced}.
In some circumstances, both types of models  share some similarities,  
for instance, the achievement of maximal asymptotic entanglement from local states~\cite{chandrashekar2012disorder,vieira2013dynamically,sen2020scaling,pires2019multiple}.
However, up to now,  only QWs with dynamic disorder in the displacement operator are able to produce hyperballistic dynamics~\cite{di2018elephant}.
For QWs with random  disorder in the coin operator, the boosting of entanglement takes place with an upper bounded-ballistic spreading ~\cite{chandrashekar2012disorder,vieira2013dynamically,zhang2022maximal}. 
Such impasse can be avoided when the temporal disorder is present in the shift operator. Specifically, recently~\cite{pires2019multiple} it was presented the first protocol -- the  generalized elephant quantum walk (gEQW)  --  that is able to exhibit both amplification of entanglement and tunable spreading from slower-than-ballistic to faster-than ballistic. However, the authors have only considered some specific and local initial states. In order to move towards  potential future applications it is necessary to answer: is the enhancement of the entanglement in the gEQW robust or only valid under the specific conditions? As will  be shown, in this work we provide a firmer conclusion: the asymptotic strengthening of the   entanglement between the  degrees of freedom in a quantum walk is robustly achieved for the protocol of the gEQW dynamics with general coin operators  and from both local and delocalized initial states.  


\section{\label{sec:model}Model}
\subsection{\label{sec:standard_dqtw}The standard coined discrete time quantum walk}


The Hilbert space of the one-dimensional coined discrete time quantum walk (DTQW) is composed of the
position space of the walker, $\mathcal{H}_p =\mbox{span}(\{\ket{x}, x \in \mathbb{Z} \})$ and the quantum coin space
$\mathcal{H}_c = \mbox{span}(\{\ket{\uparrow},\ket{\downarrow}\})$
so that the total Hilbert space of the walker is $\mathcal{H} = \mathcal{H}_p \otimes \mathcal{H}_c$ 
\cite{kempe2003quantum, venegas2012quantum, reitzner2011quantum}. Consequently, the state of the walker reads
    \begin{equation}
        \ket{\psi (t)} = \sum_{x = -\infty}^{\infty} \ket{x}(c_{\uparrow}(x, t)\ket{\uparrow} + 
        c_{\downarrow}(x, t)\ket{\downarrow})\mbox{ ,}
        \label{eq:walker_state}
    \end{equation}
whose density operator is

    \begin{equation}
        \rho(t) = \sum_{x,x'} \ketbra{x}{x'}\otimes
        \begin{pmatrix}
            c_{\uparrow}(x, t)c_{\uparrow}^{\ast}(x', t) & c_{\uparrow}(x, t)c_{\downarrow}^\ast(x', t)\\
            \\
            c_{\downarrow}(x, t)c_{\uparrow}^\ast(x', t) & c_{\downarrow}(x, t)c_{\downarrow}^\ast(x', t)
        \end{pmatrix}\mbox{ ,}
    \end{equation}
where $c_{\uparrow,\downarrow}^\ast$ denotes the complex conjugate of $c_{\uparrow,\downarrow}$.

The unitary evolution of a coined DTQW involves the action of two operators. Explicitly, we have one operator for the quantum coin, the 
\emph{coin toss operator} and the \emph{shift operator}, respectively.
The purpose of the coin toss operation is to put
the coin state in a superposition of its possible states -- a quantum analogy of the coin tossing in a classical
random walk. In the one-dimensional DTQW, the coin toss operator is a two-by-two unitary matrix, given in its most general
form by

    \begin{equation}
        C_2 =   \begin{pmatrix}
                    \cos\theta & \sin\theta e^{i\beta} \\
                    \sin\theta e^{i\gamma} & -\cos\theta e^{i(\gamma + \beta)}
                \end{pmatrix}\mbox{ ,}
        \label{eq:u2_coin}
    \end{equation}
with $\theta = \pi/4, \beta = \gamma = 0$ yielding the Hadamard operator, i.e 
    \begin{equation}
        C_2 = H = \frac{1}{\sqrt{2}}\begin{pmatrix}
                                      1 & 1 \\
                                      1 & -1 
                                  \end{pmatrix}\mbox{ .}
        \label{eq:hadamard_coin}  
    \end{equation}
In this work, we also employ 
the \emph{Kempe coin} \cite{kempe2003quantum} a class of non-hermitian coin toss operators where $\beta = \gamma = \pi/2$
    \begin{equation}
        C_2 = C_k (\theta) = \begin{pmatrix}
                \cos\theta & i\sin\theta \\
                i\sin\theta & \cos\theta  
            \end{pmatrix}
        \label{eq:kemp_coin}\mbox{.}
    \end{equation}

Regarding the shift operator, it updates the position state of the walker accordingly with its coin state.
On the one-dimensional lattice

    \begin{equation}
        S = \sum_{x = -\infty}^{\infty} \ketbra{x+1}{x}\otimes\ketbra{\uparrow}{\uparrow}
            + \ketbra{x-1}{x}\otimes\ketbra{\downarrow}{\downarrow}\mbox{ .}
    \end{equation}
Notice that the shift operator associates the up (down) state with a displacement to the right (left). 

The discrete time quantum walk one-step unitary operator is,

    \begin{equation}
        U = S(\mathbb{I}_p \otimes C)\mbox{ ,}
        \label{eq:unitary_op} 
    \end{equation}
where $\mathbb{I}_p$ is the identity operator in the position space. A recursive equation for the coefficients of the state of the walker is found if we apply the unitary operator Eq. (\ref{eq:unitary_op}) on Eq. (\ref{eq:walker_state}),
leading to
\begin{eqnarray}
    c_{\uparrow}(x, t) &=& \cos\theta c_{\uparrow}^{x-1}(t -1) + \sin\theta e^{i\beta}c_{\downarrow}^{x-1}(t -1), \nonumber \\
    & &\label{eq:recursive_rel} \\
    c_{\downarrow}(x, t) &=& \sin\theta e^{i\gamma}c_{\uparrow}^{x + 1}(t -1) - 
                         \cos\theta e^{i(\beta+\gamma)}c_{\downarrow}^{x + 1}(t -1)\mbox{ ,} \nonumber
\end{eqnarray}
where we have shortened $c_{\uparrow,\downarrow}(x\pm 1,t -1)$ to $c_{\uparrow,\downarrow}^{x\pm 1} (t - 1)$.
Recursive relations~(\ref{eq:recursive_rel}) are important tools for the analytical derivation of the asymptotic properties of the quantum walk and are quite useful to perform numerical simulations as well, as occurs in the Schr\"odinger approach method~\cite{ambainis2003quantum}.

One important point one must be aware of is that, in comparison with the classical random walk, the asymptotic properties are both initial coin-state and coin operator dependent. In other words, the outcome of a given implementation depends on the initial setting. Nonetheless, one of the remarkable facts about quantum walks is that the asymptotic behavior of the standard deviation of the position grows linearly -- i.e., ballistically --  with time, $\sigma \approx t$, whereas in the random walk it goes as $\sigma \approx t^{1/2}$ \cite{venegas2012quantum, reitzner2011quantum}. That represents a quadratic gain in the diffusion rate over the classical random walk and it is solely due to the fact that a quantum walk makes use of the superposition principle.
Another fact is that we can generate entanglement between the coin and the position degrees of freedom, a genuinely quantum feature of quantum walks, which in some cases can get to its maximal value.

\subsection{\label{sec:gEQW}The generalized elephant quantum walk}

The quantum walk model we consider consists of a noisy unitary evolution of a DTQW on a lattice where the step sizes are randomly chosen, a model devised in Ref.~\cite{pires2019multiple}, called the generalized elephant quantum walk (gEQW).
This model is an extension of the elephant quantum walk~\cite{di2018elephant}, a quantum walk inspired on the classical non-Markovian elephant random walk where the walker remember its previous steps, giving a variety of diffusion characteristics~\cite{schutz2004elephants}. Here, the 1-D shift operator reads
    \begin{equation}
        S_{t} = \sum_{x = - \infty}^{\infty} \ketbra{x + \Delta_t}{x} \otimes \ketbra{\uparrow}{\uparrow}
        + \ketbra{x-\Delta_t}{x}\otimes \ketbra{\downarrow}{\downarrow}\mbox{ ,}
    \end{equation}
where $S_t$ is the shift operator at time step $t$ and $\Delta_t$ is the step size chosen in the same time 
instant. In this way, for every time instant we will have a random unitary operator Eq. (\ref{eq:unitary_op}), with $S_t$ in place of $S$, using the coin operators as Eq. (\ref{eq:hadamard_coin}) and 
Eq. (\ref{eq:kemp_coin}) in the same manner.

We can interpret a random unitary evolution as a noisy open evolution where one observes 
at each time step which unitary operator was selected by the environment. Let $\mathcal{H}_E$ be the Hilbert 
space of the environment and $\{\ket{\Delta_j},\; j = 1,\dots,t\}$ be a spanning set of it. Given that 
the system-environment state $\rho_{S,E} \in \mathcal{H}_S \otimes \mathcal{H}_E$ is closed, it evolves through 
a unitary operator $U$

    \begin{equation}
        U = \sum_{j = 1,\dots,t} U_j \otimes \ketbra{\Delta_j}{\Delta_j}
        \mbox{ .}
    \end{equation}
Supposing that the environment state is given by, $\ket{\psi_E(t)} = \sum_{j = 1,\dots,t}\sqrt{p(\Delta_j)}\ket{\Delta_j}$,
then, the total state evolves as

    \begin{equation}
        \rho_{S,E}(t+1) = \sum_{j,j'} \sqrt{p(\Delta_j)}\sqrt{p(\Delta_j')}U_j\rho_S(t)U_{j'}^{\dagger} \otimes \ketbra{\Delta_j}{\Delta_j'}
        \mbox{ ,}
    \end{equation}
Performing a projective measurement $P_t = \ketbra{\Delta_t}{\Delta_t}$ and a partial trace over the environment degree of freedom
we obtain the following unnormalized system state 

    \begin{equation}
        \rho_S(t+1) = p(\Delta_t)U_t\rho_S(t)U_t^{\dagger}\mbox{ ,}
    \end{equation}
whose norm $p(\Delta_t)$ is the probability that the state $U_t\rho_S(t)U_t^\dagger$ was selected from the statistical mixture
$\sum_{j} p(\Delta_j)U_j \rho_S(t)U_j^{\dagger}$.

\begin{figure}[!ht]
        \centering
        \includegraphics[scale = 0.325]{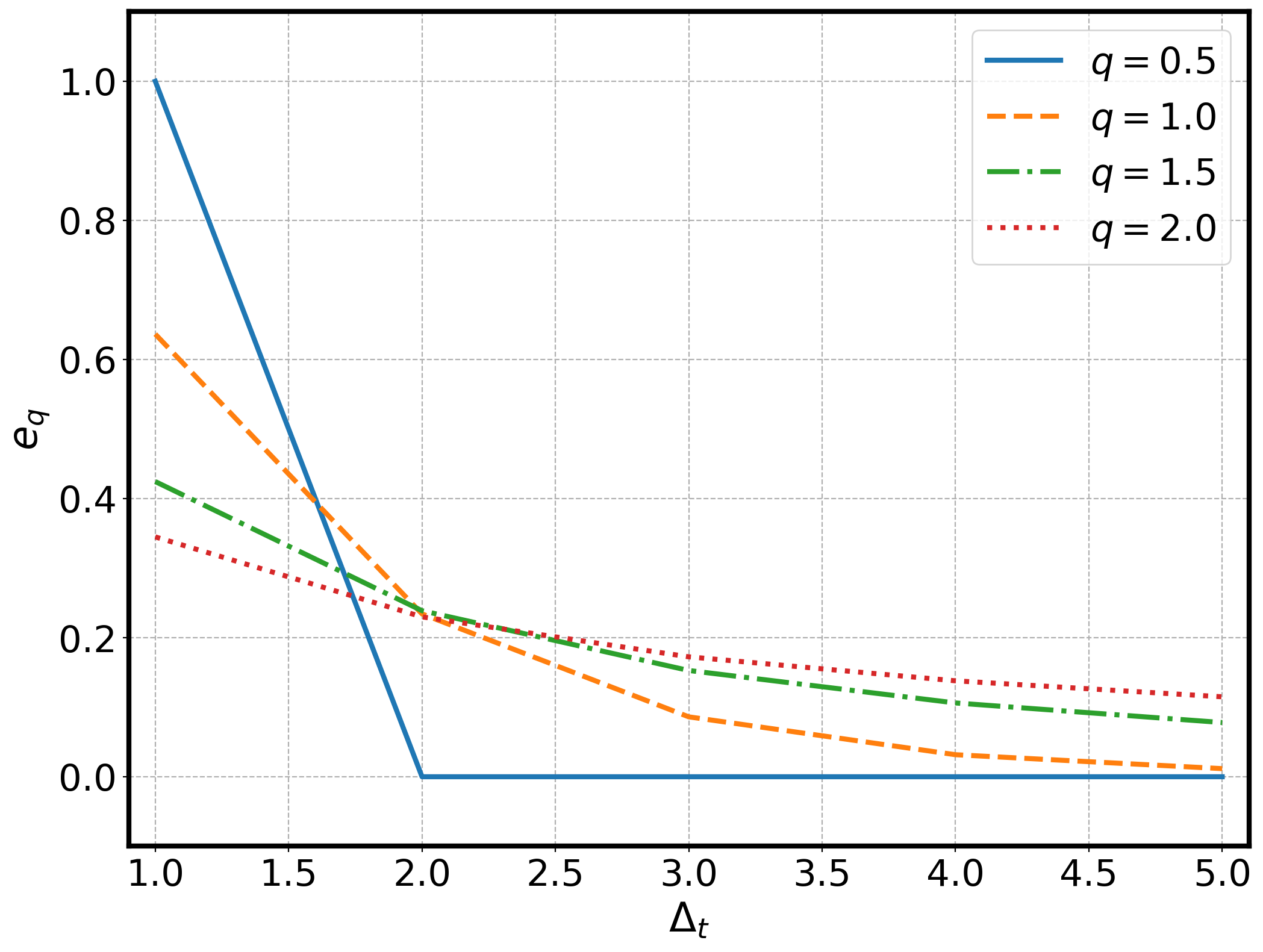}
        \caption{q-Exponential probability distribution for some values of $q$.}
        \label{fig:qExp}
    \end{figure}

The probability distribution used for choosing the steps sizes is a discretized version of the $q$-exponential
distribution \cite{tsallis2009nonadditive} (see Fig.~\ref{fig:qExp})

    \begin{equation}
        \mbox{Pr}(\Delta_t) = e_{q}(\Delta_t) \equiv \tau_t[1 - (1 - q)\Delta_t]^{1/1-q}\mbox{,}
        \label{eq:qexp}
    \end{equation}
with $\Delta_t \in [1,2,\dots,t]$, $\tau_t$ being a time-dependent normalization 
factor and with support given by
    \begin{equation}
        \mbox{supp}(e_q(x)) =   \begin{cases}
                                    [0, \frac{1}{1-q})\mbox{, }q \le 1\\
                                    [0 , \infty)\mbox{, }q > 1\mbox{ .}
                                \end{cases}
        \label{eq:supp_qexp}
    \end{equation}
    
It is worth noting some limiting cases of the $q$-exponential probability distribution.
For instance, when we set $q = 1/2$, it reads
    \begin{equation}
       \displaystyle e_{1/2}(\Delta_t) = 
       \tau_t\left(1 - \frac{\Delta_t}{2}\right)^2\mbox{.}
    \end{equation}
Looking at Eq. (\ref{eq:supp_qexp}) we see that $\Delta_t \le 2$, 
where $e_{1/2}(\Delta_t = 1) = 1$ and $e_{1/2}(\Delta_t = 2) = 0$.
Therefore, with $q = 1/2$ only unit step sizes are possible, matching with the standard 
DTQW. On the other hand, when $q = 1$, we get a decreasing exponential in the step sizes
    \begin{equation}
        \lim_{q \rightarrow 1}e_{q}(\Delta_t) = \tau_t e^{-\Delta_t}\mbox{.}
    \end{equation}
Lastly, taking the limit of $q$ going to infinity we get the uniform distribution

    \begin{equation}
        \lim_{q \rightarrow \infty}e_{q}(\Delta_t) = \frac{\tau_t}{t}\mbox{ ,}
    \end{equation}
characterizing the Elephant Quantum Walk (EQW). Looking through the open evolution perspective, 
the $q$-exponential distribution give us a versatile way to model a myriad of quantum walks ranging from the deterministic evolution, 
corresponding to the standard DTQW, to the completely random one, namely the EQW.


In order to compare the degree of dispersion of a quantum walker it was analyzed the asymptotic limit of the variance of the position \cite{pires2019multiple}. It is expected that its limiting behavior obeys
    \begin{equation}
        \mbox{Var}_X(t)  = \sigma^{2}_{X} \approx t^{\alpha}\mbox{, }\; t \gg 1 \mbox{,}
    \end{equation}
where $\alpha$ is called \emph{diffusion exponent}. By taking the logarithm of the position variance graph it is possible to estimate the diffusion exponent and compare the quantum walks dispersion for different values of $q$. We must recall that the evolution of this type of DTQW is random, in the sense that for a given time instant the unitary operator can be different 
for different runs of the quantum walk. Consequently, the most appropriate is to consider the average diffusion exponent.
    
\begin{figure}[!ht]
            \centering
            \includegraphics[scale=0.315]{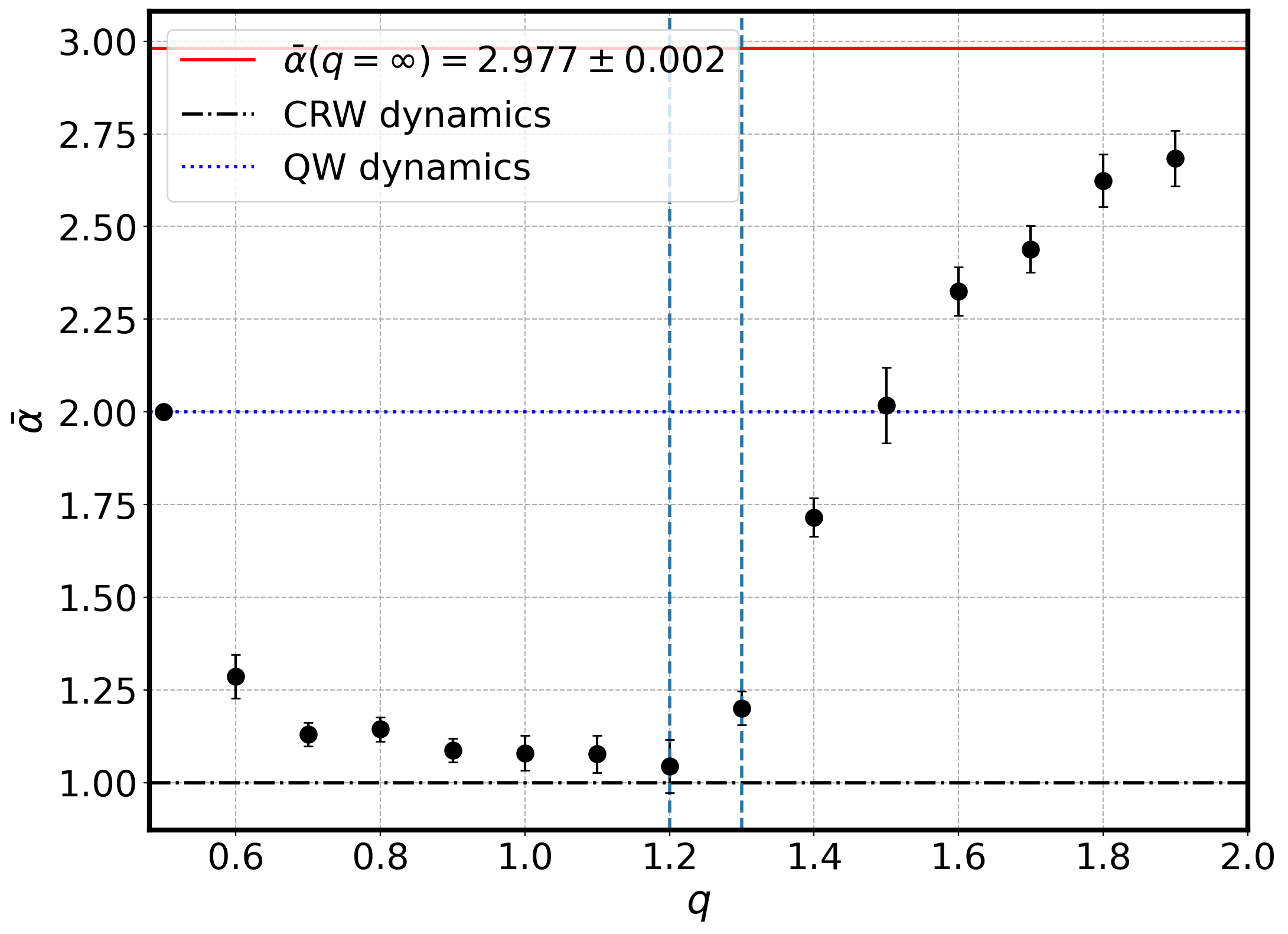}
            \caption{(Color online) Mean diffusion exponent as a function of the $q$ parameter 
            considering the quasi-stationary part of the evolution using the Kempe coin Eq. (\ref{eq:kemp_coin})
            with $\theta = \pi/4$ and the initial quantum walker state 
            $\ket{\psi(0)} = \ket{0}\otimes (\ket{\uparrow} + \ket{\downarrow})/\sqrt{2}$. The dashed line and dash-dotted line
            indicates the standard DTQW diffusion exponent $\alpha = 2$ and the classical random walk diffusion $\alpha = 1$, respectively.
            The vertical lines indicate the range of $q$ above which the QW has an increasingly faster spreading. 
            The red line give us the value of EQW's diffusion exponent $\bar{\alpha} = 2.977 \pm 0.002$.}
            \label{fig:mean_alpha_x_q}
    \end{figure}

In Fig.~\ref{fig:mean_alpha_x_q}, we have the mean diffusion exponent as a function of $q$, $\bar{\alpha}(q)$, in the range $[0.5,1.9]$, considering only the quasi-stationary part of the quantum walks evolution. 
We also considered the value of the mean diffusion exponent for $q = \infty$. The behavior of the curve defined by the data points is similar to that obtained in Ref.~\cite{pires2019multiple}. For $q > 1.3$, the diffusion exponent starts increasing and it reaches its asymptotic hyper-ballistic limit, $\bar{\alpha} = 3$. Bearing in mind that $q = \infty$ corresponds to the uniform distribution case, i.e., strong randomness in the step sizes leads the quantum walk to a hyper-ballistic regime whereas a weak randomness can lead it either to the standard DTQW ballistic regime or to the random walk diffusion.

In Ref.~\cite{pires2019multiple}, it was shown that considering some values of $q$ different to $q = 1/2$ and employing the von Neumann entropy as purity quantifier of the coin state, the gEQW produces maximally entangled coin states when using the Kempe coin operator Eq.~(\ref{eq:kemp_coin}) with $\theta = \pi/4$ (see Fig.~5 in Ref.~\cite{pires2019multiple}).
The questions we address are as follows: what if that result depends on the initial state of the coin? Is coin operator dependent as well? How does the average entropy change when we change the $q$-exponential distribution, i.e., the amount of disorder?

In the next section, we analyze the entanglement between the coin and position degrees as a function of the coin initial state
and coin operator parameters in the gEQW, taking initially localized and Gaussian delocalized walker states. After it, we investigate how the generalize elephant quantum walk goes to the quasi-stationary regime.
    
\section{\label{sec:results}Results and discussion}

\subsection{\label{sec:res_entang}Coin entanglement entropy in the gEQW}

    \begin{figure*}[!ht]
        \centering
        \includegraphics[scale = 0.29]{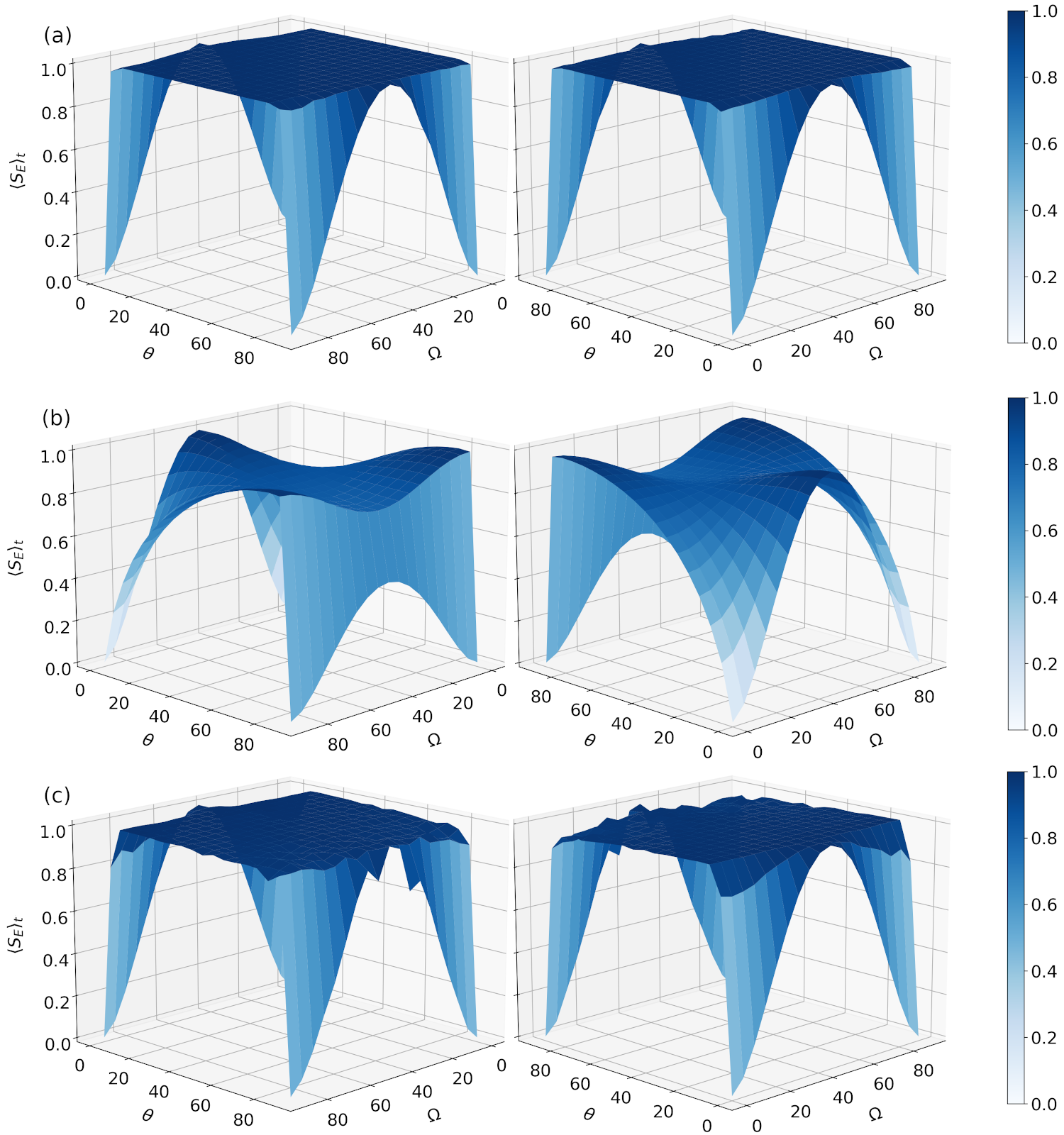}
        \caption{Time average entanglement entropy as a function of the $\theta (\deg)$ parameter in the Kempe coin operator Eq.(\ref{eq:kemp_coin}) and the Bloch
        polar angle $\Omega (\deg)$ of the coin initial state Eq.(\ref{eq:coin_instate}) in the generalized elephant quantum walk with $q = \infty$ (a),
        $q = 1/2$ (b) and $q = 1$ (c). The values of $\theta$ and $\Omega$
        considered to generate this graph were taken in intervals of $5^{\circ}$, from $0^{\circ}$ to $90^{\circ}$.
        The average entanglement entropy for every point was obtained considering only the quasi-stationary part of the entropy time evolution and
        an initially localized walker was considered in all simulations. }
        \label{fig:mentang}
    \end{figure*}
    
The von Neumann entropy of a quantum state is defined as~\cite{horodecki2009quantum}

    \begin{equation}
        S_E \equiv  -\mbox{tr}(\rho \log_2\rho) = - \sum_{i} \lambda_{i}\log_2\lambda_i\mbox{ ,}
    \end{equation}
with $\lambda_i$ being the eigenvalues of the density matrix, $\rho$. We choose to consider the von Neumann entropy of the coin density matrix because it is a two-level system and the simplest part of the position-coin bipartition.

The coin density matrix is given by
    \begin{equation}
        \rho_c(t) \equiv \mbox{tr}_{x}(\rho(t)) = \begin{pmatrix}
                                    A(t) & B(t) \\
                                    B^*(t) & C(t)
                                \end{pmatrix}\mbox{ ,}
        \label{eq:coin_density}
    \end{equation}
where $A(t) = \sum_{x} |c_{\uparrow}(x, t)|^2$, $B(t) = \sum_x c_{\uparrow}(x, t)c_{\downarrow}^\ast(x, t)$
and $C(t) = \sum_x |c_{\downarrow}(x, t)|^2$. In order to find the coin state coefficients in the gEQW, we have to change the recursive relations of Eq.~(\ref{eq:recursive_rel}) to
\begin{align}
    c_{\uparrow}^{x}(t) &= \cos\theta c_{\uparrow}^{x-\Delta_t}(t -1) + \sin\theta e^{i\beta}c_{\downarrow}^{x-\Delta_t}(t -1)\\
    c_{\downarrow}^{x}(t) &= \sin\theta e^{i\gamma}c_{\uparrow}^{x + \Delta_t}(t -1) - 
                         \cos\theta e^{i(\beta+\gamma)}c_{\downarrow}^{x + \Delta_t}(t -1)\mbox{.}
    \label{eq:geqw_rec_rel}
\end{align}
Accordingly, the von Neumann entropy of the coin is given by
    \begin{align}
        S_E &= -\lambda_+\log_2\lambda_+ - \lambda_-\log_2\lambda_-\mbox{ , with} \\
        \lambda_{\pm} &= \frac{1}{2}\left( 1 \pm \sqrt{(1 - 4(AC - |B|^2)}\right) \mbox{ ,}
    \end{align}
where we used the fact that $\mbox{tr}(\rho_c(t)) = 1 \rightarrow A(t) + C(t) = 1,\; \forall t$.
%
    
As a way to study the entanglement generation as a function of the initial parameters, throughout this work we have considered the time average entanglement entropy taking only the quasi-stationary part of the entanglement evolution. It is expected that, after an initial increase, the entanglement entropy reaches, at least, an average constant value~\cite{vieira2013dynamically,vieira2014entangling}. Given the quasi-stationary regime varies with the type of quantum walk, initial state and coin operator, it was determined individually for each evolution analyzed here by looking at the hole von Neumann entropy time evolution.

First, we study the time-averaged entanglement entropy
as a function of the $\theta$ parameter in the Kempe coin operator, Eq. (\ref{eq:kemp_coin}), and the polar angle $\Omega$
on the Bloch sphere of the coin initial state,
    \begin{equation}
        \ket{\psi_c (0)} = \cos\left(\frac{\Omega}{2}\right)\ket{\uparrow} + e^{i\phi/2}\sin\left(\frac{\Omega}{2}\right) \ket{\downarrow}\mbox{ ,}
        \label{eq:coin_instate}
    \end{equation}
with $q \rightarrow \infty$, the elephant quantum walk scenario, and $\phi$ set to zero. 
From Fig.~\ref{fig:mentang}(a), we see the blue plateau indicates that the average entanglement entropy reaches its maximum value for almost all initial states and Kempe coin operators something that does not happen when we have the standard discrete time quantum walk (see e.g. Fig.~\ref{fig:mentang}(b)). 
Using a different value of $q$, e.g. $q = 1$, a similar result is obtained Fig.~\ref{fig:mentang}(c).
These results, together with that obtained in Ref.~\cite{pires2019multiple}, indicate the generalized elephant quantum walk has the potential to generate maximally entangled coin states for almost all initial coin parameters and Kempe coin operators, considering $q \ne 1/2$ and an initially localized walker state.

\begin{figure}[!ht]
        \centering
        \includegraphics[scale = 0.315]{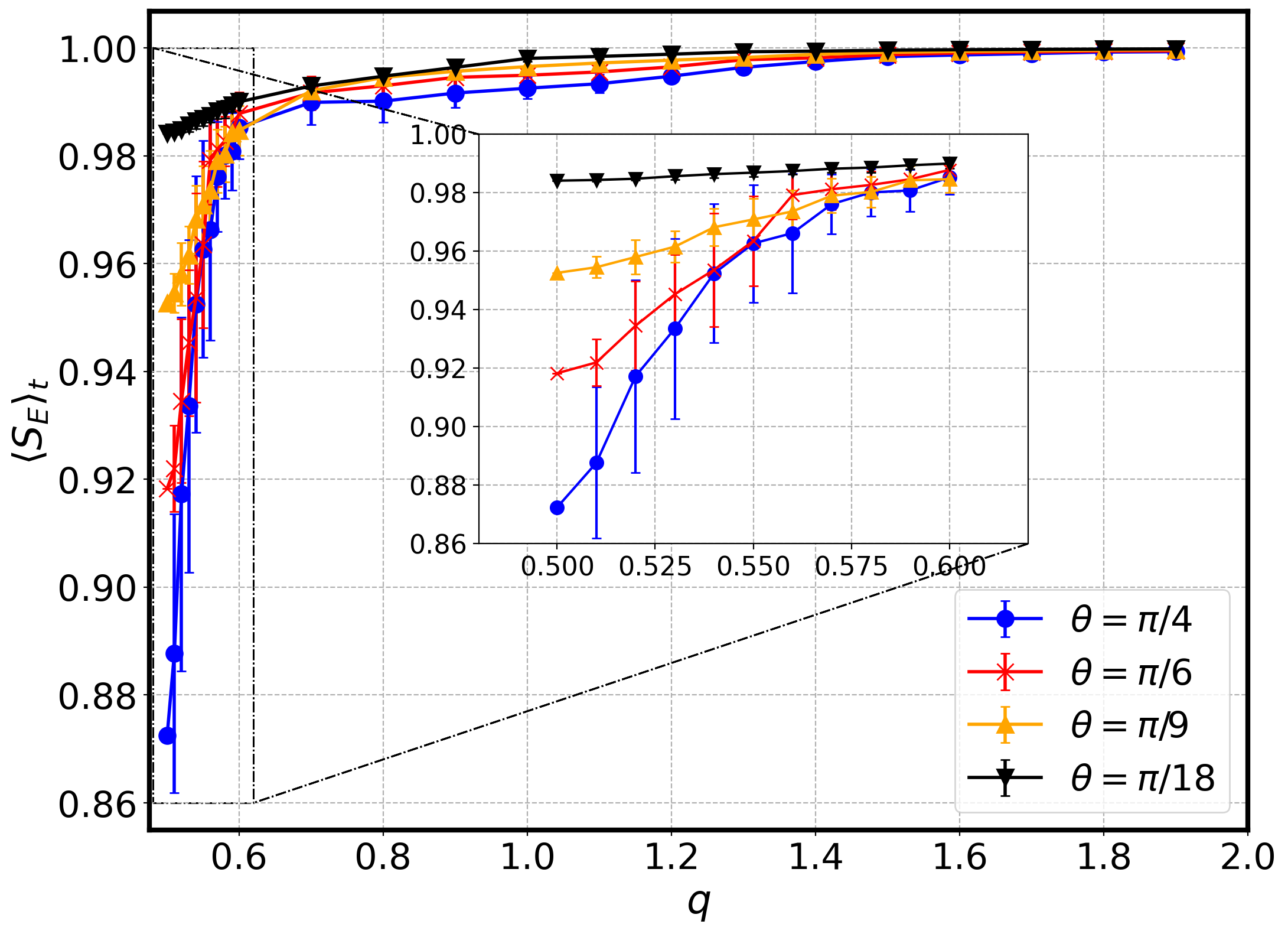}
        \caption{(Color online) Time average entanglement entropy as a function of $q$ in the q-exponential distribution 
        Eq. (\ref{eq:qexp}) in the gEQW for 
        different values of $\theta$ in the Kempe coin Eq. (\ref{eq:kemp_coin}). The data points
        were obtained through the average of $50$ simulations each and the error bars indicate the standard deviation of the
        points. In all simulations the initial state was $\ket{0}\otimes (\ket{\uparrow} + \ket{\downarrow})/\sqrt{2}$,
        i.e. $\Omega = \pi/2$ and $\phi = 0$.}
        \label{fig:me_x_q}
    \end{figure}

In Fig.~\ref{fig:me_x_q}, we analyze the time-averaged entanglement entropy of the coin system as a function of the $q$ parameter for some values of $\theta$ in the Kempe coin operator. From it, we observe that the entanglement entropy increases very fast in the interval $[0.5, 0.6]$ and goes asymptotically to $\langle S_E \rangle_t = 1$ as $q \rightarrow \infty$. Going back to the $q$-exponential function, changing from $q = 0.5$ to $q = 0.6$ we only soar the probability of having steps of size equal to $2$ from $0$ to approximately $6\%$. However, we have a substantial increase in the average entanglement, going from $0.8724$ to $0.9852$ for $\theta = \pi/4$, and a moderate increase for $\theta = \pi/6$, from $0.9183$ to $0.9878$. With $\theta = \pi/18$, as the average entanglement with $q = 0.5$ is already significant (as can be seen in Fig.~\ref{fig:mentang}(b)), the increase is also small. 
In the long time limit, changing the parameter $q$ to one the probability of unit step sizes is approximately $63\%$, of step sizes equal to two approximately $23\%$, while of steps of sizes equal to three $9\%$, but for all $\theta$ we already have an almost fully entangled state of $\langle S_E \rangle_t \approx 0.99$. 
Consequently, considering an initially localized walker and the Kempe coin, we say that by allowing steps $\Delta_t = 2$ with a small probability, we enhance in the generation of entanglement between the coin and position subsystems, not being necessary a strong randomness in the step sizes, and with a probability of approximately $9\%$ of $\Delta_t = 3$ the time-averaged coin von Neumann entropy almost reaches its maximum value.

    \begin{figure}[!ht]
        \centering
        \includegraphics[scale = 0.31]{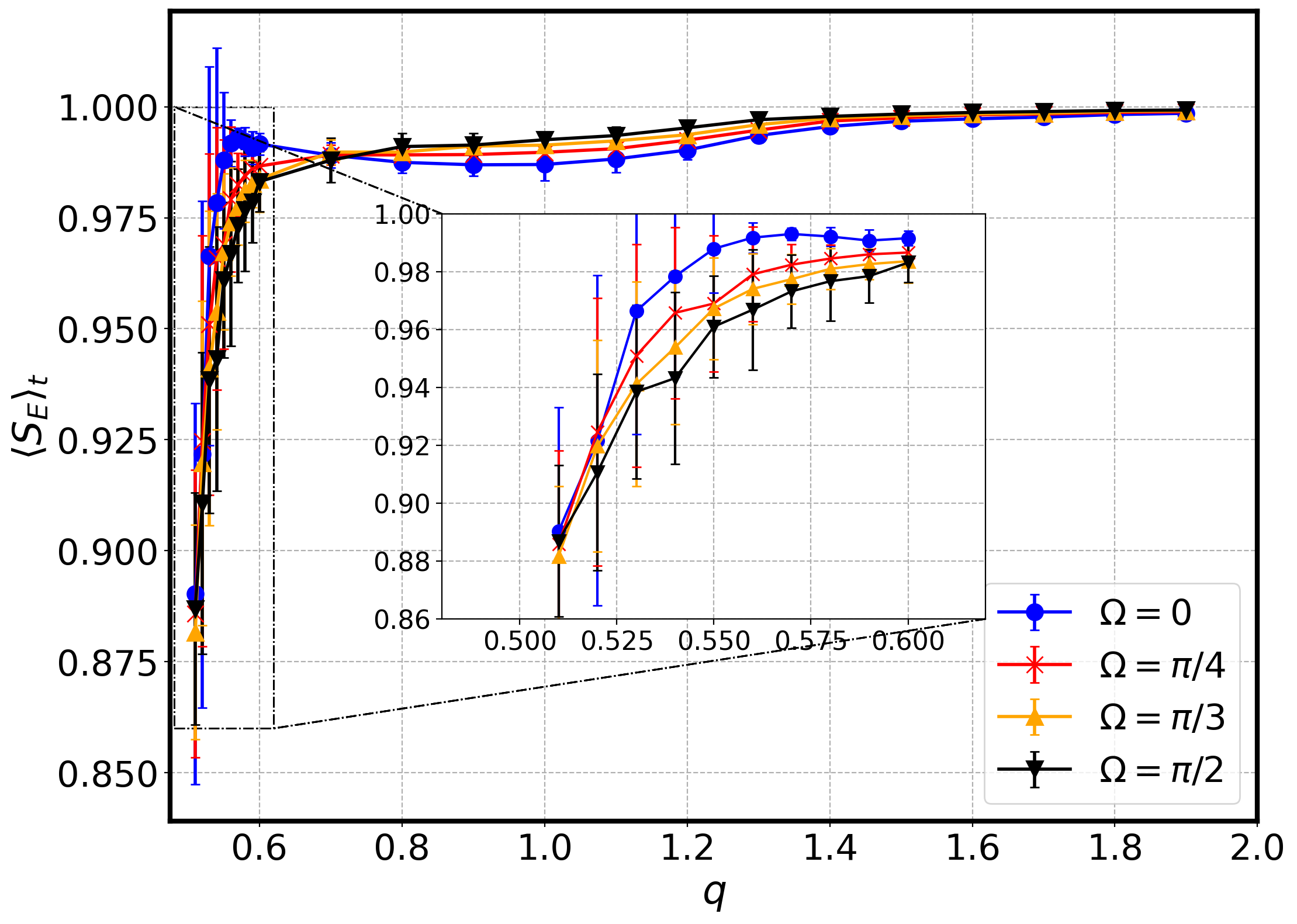}
        \caption{(Color online) Time average entanglement entropy as a function of $q$ in the q-exponential distribution
        Eq. (\ref{eq:qexp}) in the gEQW with the Kempe coin operator Eq. (\ref{eq:kemp_coin}) with $\theta = \pi/4$
        for different values of $\Omega$ using $\phi = 0$ for all of them.
        The data points were obtained through the average of $50$ simulations
        each and the error bars indicate the standard deviation of the points. In all simulations the localized
        initial state was used.}
        \label{fig:me_x_q_beta}
    \end{figure}
    
We have studied how the time average entanglement entropy of the gEQW for $\theta = \pi/4$ in Eq. (\ref{eq:kemp_coin})
behaves as we change the coin initial state Bloch polar angle $\Omega$. In Fig.~\ref{fig:me_x_q_beta} by varying $\Omega$ we are able to control the increase of the entanglement entropy for $q$ in the interval $[0.5, 0.6]$, where for $\Omega = 0$ the greatest rate is found. For $q > 0.6$, the entanglement entropy decreases thus swapping the proportionality relation between $\Omega $ and the entanglement entropy by increasing parameter $q$, which only converges with the other curves at $q = 1.6$, approximately.

\begin{figure}[!ht]
        \centering
        \includegraphics[scale = 0.315]{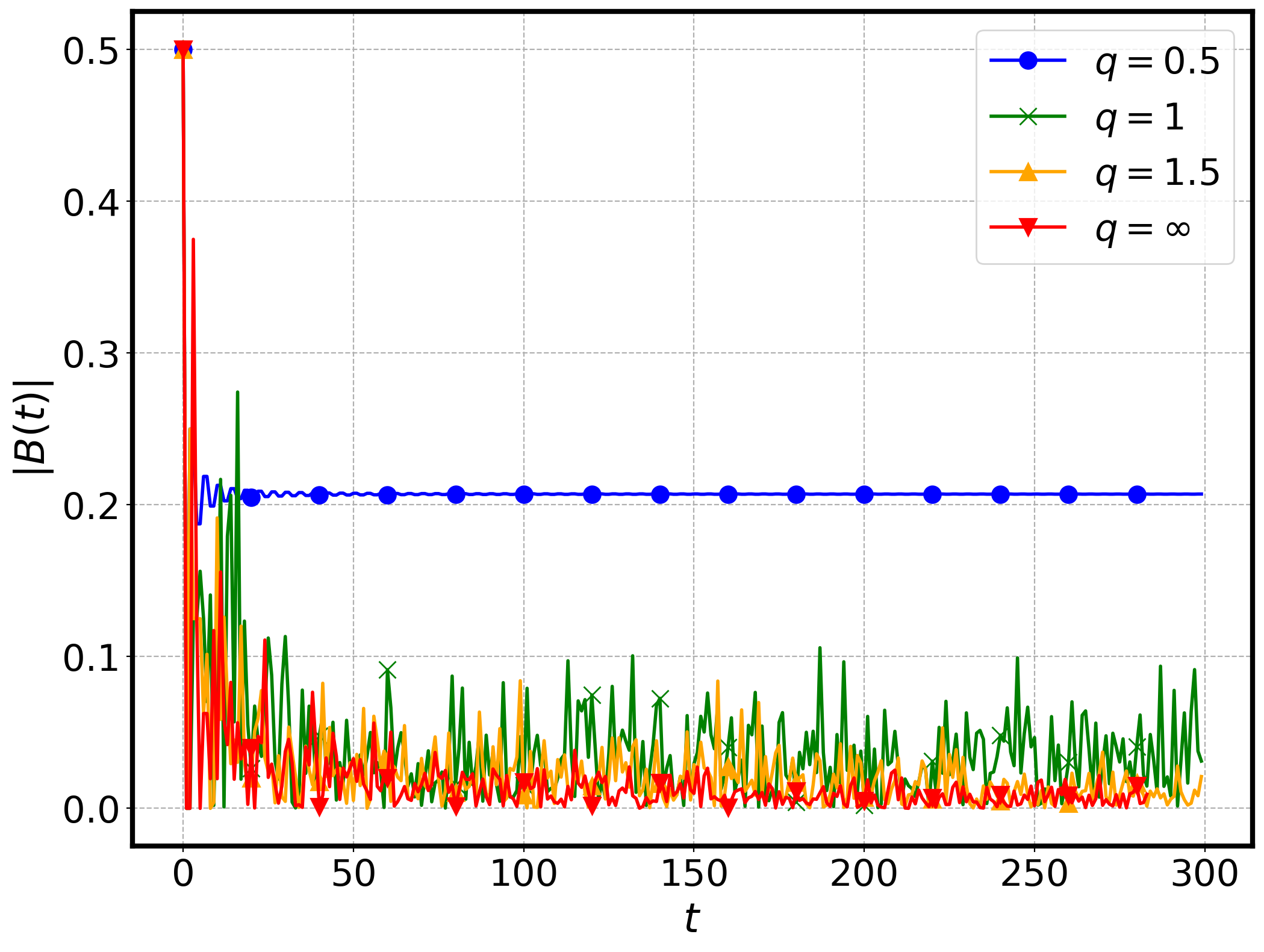}
        \caption{(Color online) Time evolution of the coin density matrix coherence absolute value for different generalized elephant quantum walks using $C_k(\pi/4)$.
        The initial state used in all simulation was the one localized on the origin and with the parameters $\phi = 0$ and $\Omega = \pi/2$ for the coin.}
        \label{fig:coherences_x_q}
    \end{figure}

One can get a physical intuition why the generalized elephant quantum walk yields highly entangled states by remembering its open quantum walk interpretation. The surrounding environment that selects which unitary evolution that the walker will go under introduces a decoherence effect in the coin evolution that can be 
revealed by the coin density coherence time evolution. Figure~\ref{fig:coherences_x_q} shows the time evolution of the absolute value of $B(t)$ in the coin density matrix Eq. (\ref{eq:coin_density}) for different $q$ values of the generalized elephant quantum walk using the balanced Kempe coin. When considering the standard DTQW $q = 0.5$, we see 
that the coherence absolute value has a decaying oscillating behavior, stabilizing into a value of approximately $0.2$. By increasing the randomness on the step sizes, it decays even faster and stabilizes into lowers values according to the degree of randomness, going to zero for the maximally random case, i.e. the EQW. This behavior is in agreement with the observed behavior of the average entanglement entropy as a function of $q$ Figs.~\ref{fig:me_x_q}~and~ \ref{fig:me_x_q_beta}.

    \begin{figure*}
        \centering
        \includegraphics[scale = 0.29]{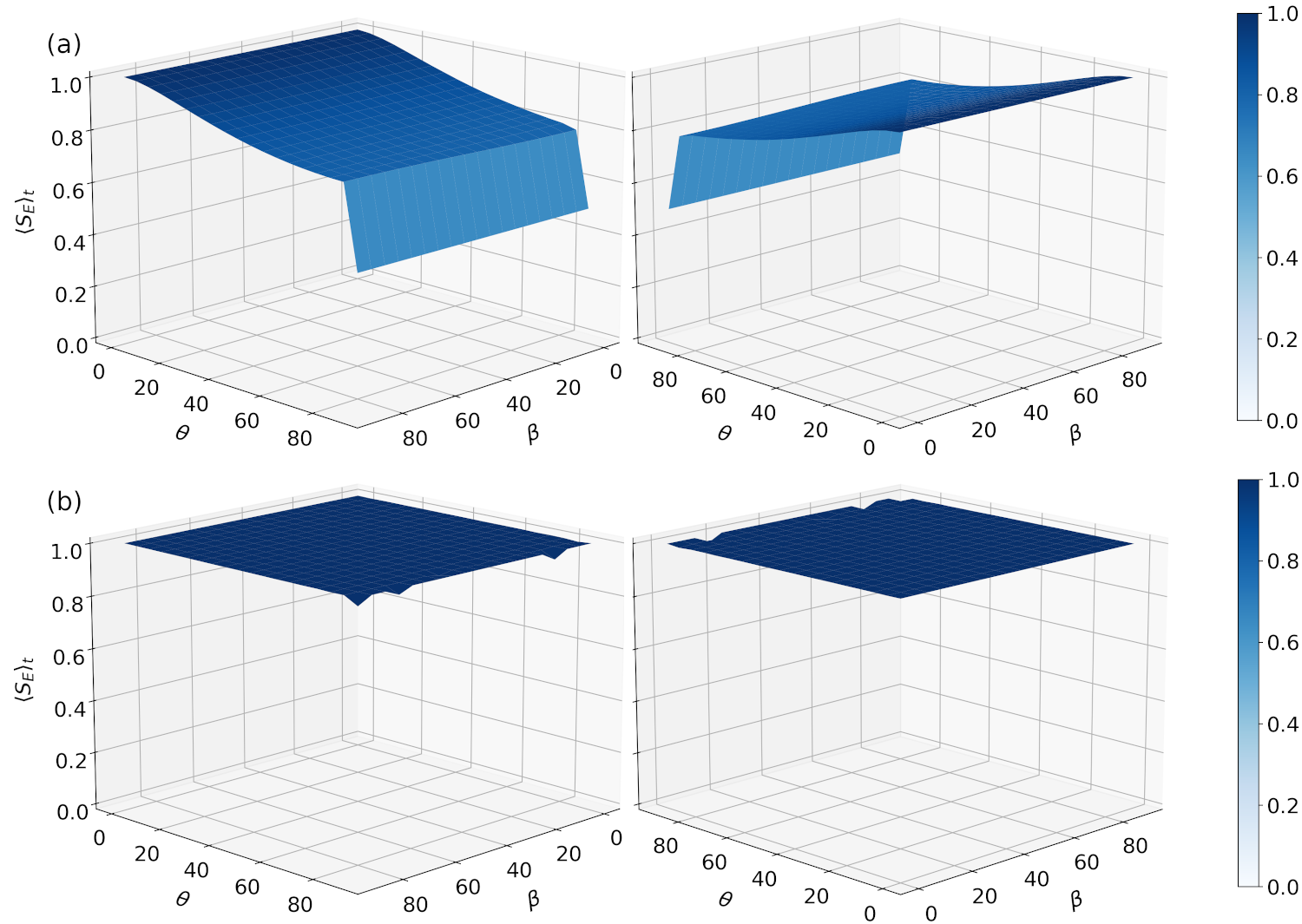}
        \caption{Time average entanglement entropy of the coin state in the generalized elephant quantum walk
        as a function of $\theta (\deg)$ and $\beta (\deg)$ in the coin operator Eq. (\ref{eq:u2_coin}). In (a) we have
        $q = 0.5$ and in (b) $q = \infty$. All simulations were done considering an initially localized walker
        state and $\Omega = \pi/2$ and $\phi = 0$ in Eq. (\ref{eq:coin_instate}).}
        \label{fig:mentang_3d_beta}
    \end{figure*} 

Next, we investigate the time-averaged entanglement entropy by modifying the phase angles of the coin operator, Eq.~(\ref{eq:u2_coin}), namely $\beta$, as depicted in the 3D plots of the mean entanglement entropy as a function of $\theta$ and $\beta$ for $q = 0.5$, Fig.~\ref{fig:mentang_3d_beta}(a), and $q \rightarrow \infty$, Fig.~\ref{fig:mentang_3d_beta}(b).

Taking an initially localized walker state and $\Omega = \pi/2$ with $\phi = 0$, we understand that by varying the phase angle $\beta$ in the standard quantum walk the time average entanglement entropy of the coin state does not changes for a given value of $\theta$, as $\langle S_E \rangle_t$ vs. $\beta$ remains constant. The same conclusion is drawn for the elephant quantum walk in Fig.~\ref{fig:mentang_3d_beta}(b). In addition, it does not matter whether we vary $\theta$ for a given value of $\beta$, because $\langle S_E\rangle_t$ vs. $\theta$ remains virtually constant. That is a strong indicative that the generalized elephant quantum walk, for $q \ne 1/2$, produces highly entangled coin states, $ S_E > 0.87$ for $ q \in (0.5, 0.6]$ , and maximally entangled coin states for $q \rightarrow \infty$, for all coin operators and coin initial states, considering an initially localized walker state.

Following the above results, we survey the time averaged coin von Neumann entropy for delocalized Gaussian walker initial states.
\subsection{Delocalized initial states}

The form of the delocalized position initial states that we considered is Gaussian

    \begin{equation}
        \ket{\psi_p(0)} = \sum_{x = -\infty}^{\infty} N e^{\frac{-x^2}{4\sigma^2}}\ket{x}\mbox{ ,}
        \label{eq:gauss_instate}
    \end{equation}
where $N$ is a normalization factor and $\sigma$ the standard deviation of the distribution.
In the standard DTQW, by using delocalized initial states the position variance only gets a polynomial form in the short time period, like Var$_x(t) = a_0 + a_1 \; t + a_2 \; t^2$. 
Regarding the coin entanglement entropy, in Ref.\cite{orthey2017asymptotic} was studied
the asymptotic coin state when one considers a Gaussian distribution for the position initial state as well, but for a Hadamard walk, Eq. (\ref{eq:hadamard_coin}). They found a relation between the coin initial state angles on the Bloch sphere that gives a maximally entangled coin state

    \begin{equation}
        \cos{\phi} = -\cot{\Omega}\mbox{ ,}
        \label{eq:orthey_rel}
    \end{equation}
when the initial position variance $\sigma \gg 1$.

    \begin{figure}[!ht]
        \centering
        \includegraphics[scale = 0.33]{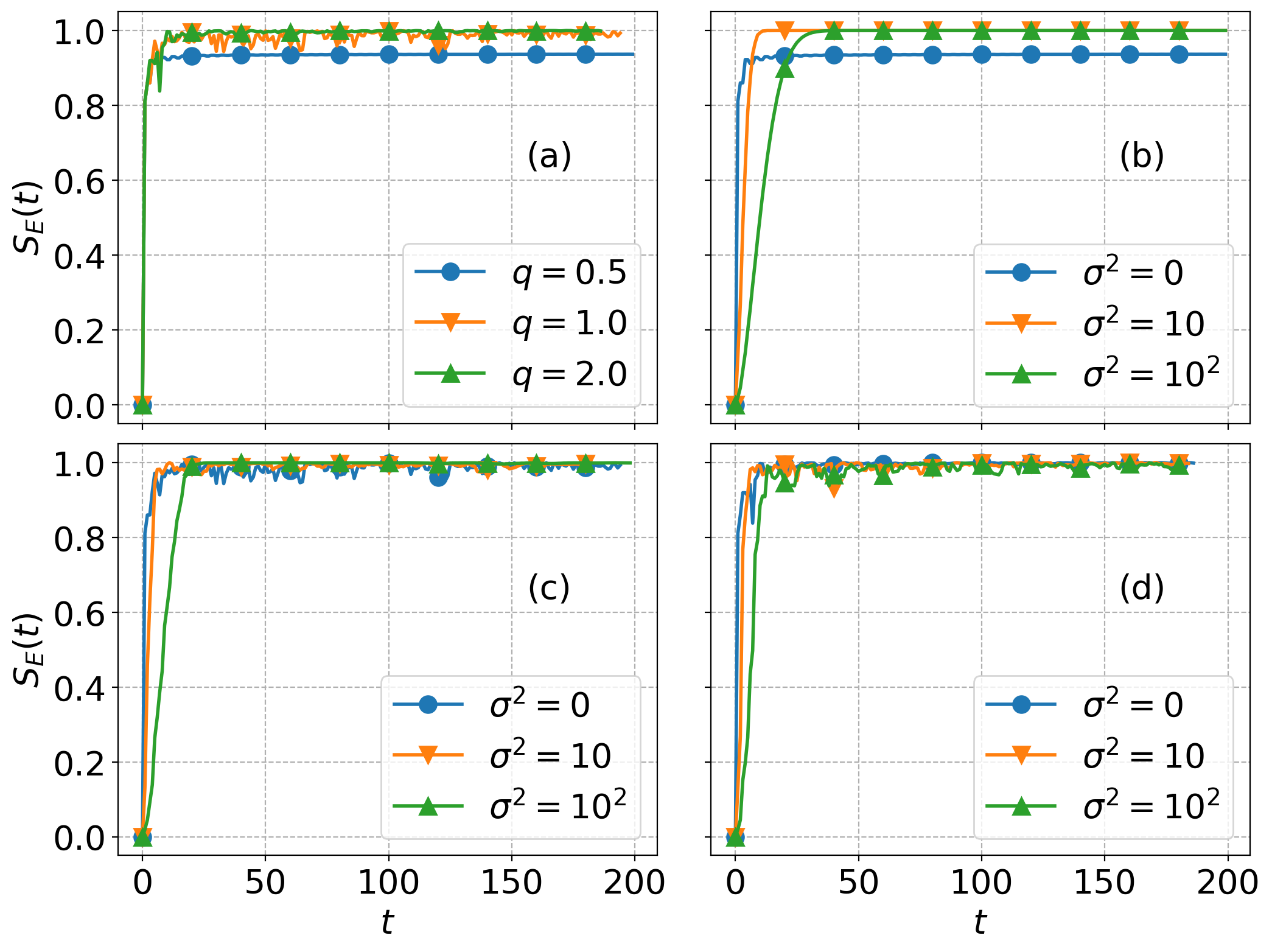}
        \caption{Time evolution of the coin entanglement entropy for (a) different values of $q$ and $\sigma^2 = 0$, 
         and different values of $\sigma$ and $q = 0.5$ (b), $q = 1$ (c), $q = 2$ (d)
        in the Hadamard Walk.
        The coin initial state used was the one following Eq. (\ref{eq:orthey_rel}) with $\Omega = \pi/3$.}
        \label{fig:Se_x_t_gauss}
    \end{figure}
    
Aiming at capture the effect on the entanglement entropy of introducing randomness in the step sizes, we have computed the time evolution of the entanglement entropy for an initial coin state with $\Omega = \pi/3$ and $\phi \approx 0.696\pi$, following Eq.~(\ref{eq:orthey_rel}) in Fig.~\ref{fig:Se_x_t_gauss}.
In the top left panel Fig.~\ref{fig:Se_x_t_gauss}(a), we see the quantum walk with random step sizes leads the entanglement entropy to the maximum value while the same does not happens for the standard DTQW where an initially localized
state is considered. That is in agreement with our previous results. However, as we change the variance of initial position Fig.~\ref{fig:Se_x_t_gauss}(b), the coin entanglement entropy gets to the maximal, reproducing previous results~\cite{orthey2017asymptotic}.
The only significant difference between the initially localized and delocalized states in the cases where we use the gEQW Fig.~\ref{fig:Se_x_t_gauss}(c,d) are in the increase rate of the entanglement entropy as a function of time, where as we increase $\sigma$ we get a slower $S_E(t)$ initial increase.
    
\begin{figure*}[!ht]
        \centering
        \includegraphics[scale = 0.29]{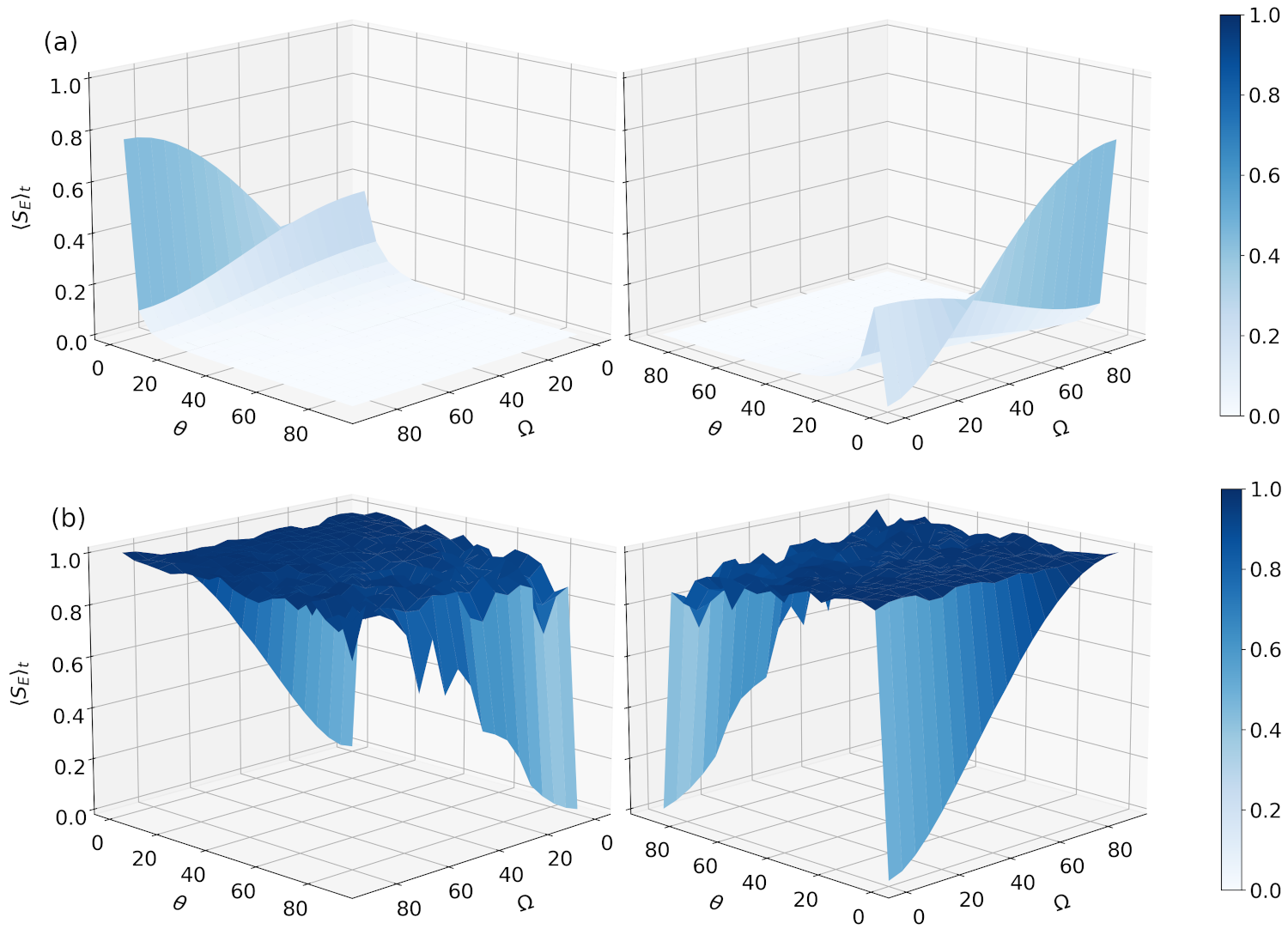}
        \caption{Time average entanglement entropy as a function of $\theta (\deg)$ in Eq. (\ref{eq:kemp_coin}) and
        $\Omega (\deg)$ in Eq. (\ref{eq:coin_instate}) for the gEQW with $q = 1/2$ (a) and $q = \infty$ (b). 
        The position initial state used was a Gaussian distribution Eq. (\ref{eq:gauss_instate}) with
        $\sigma^2 = 10^3$ for both plots.}
        \label{fig:me3d_gauss}
    \end{figure*}

The following 3D plot shows the time-averaged entanglement entropy as a function of the Kempe coin operator parameter and the coin initial Bloch polar angle in the standard DTQW Fig.~\ref{fig:me3d_gauss}(a), 
where we have considered a Gaussian initial state with $\sigma^2 = 10^3$. 
It is visible that the average coin entanglement entropy has lowered for all $\{\theta, \Omega\}$ pairs, with the maximum value obtained when we set $\theta = 0^{\circ}$ and $\Omega  = 90^{\circ}$. For almost all pairs with $\theta > 20^\circ$ the coin entanglement entropy
reaches its minimal value. In other words, the coin-position system  
is a separable one, something that happens only for a few points in the localized initial state case (see Fig.~\ref{fig:mentang}).
Therefore, in the standard DTQW the introduction of highly delocalized walker initial states drastically affects the asymptotic entanglement.


Figure~\ref{fig:me3d_gauss}(b) shows us the 3D plot of the elephant quantum walk case. Therein, we note that for almost all pairs the average entanglement entropy is still close to its supreme, but with more oscillations around it. Furthermore, the behavior of the surface on the regions where $\theta \approx 0^\circ$ or 
$\theta \approx 90^\circ$ has significantly changed, with a decrease of $\langle S_E \rangle_t$ to $0.8$ as $\theta$ goes to $90^\circ$ and $\Omega$ goes to $0^\circ$. This scenario indicates us that the coin entanglement entropy in the elephant quantum walk using the Kempe coin operator is robust against the use of highly delocalized walker initial states for a significant part of the set of possible $\{\theta, \Omega\}$ pairs, while this does not happen in the standard quantum walk.

\begin{figure}[!ht]
        \centering
        \includegraphics[scale = 0.315]{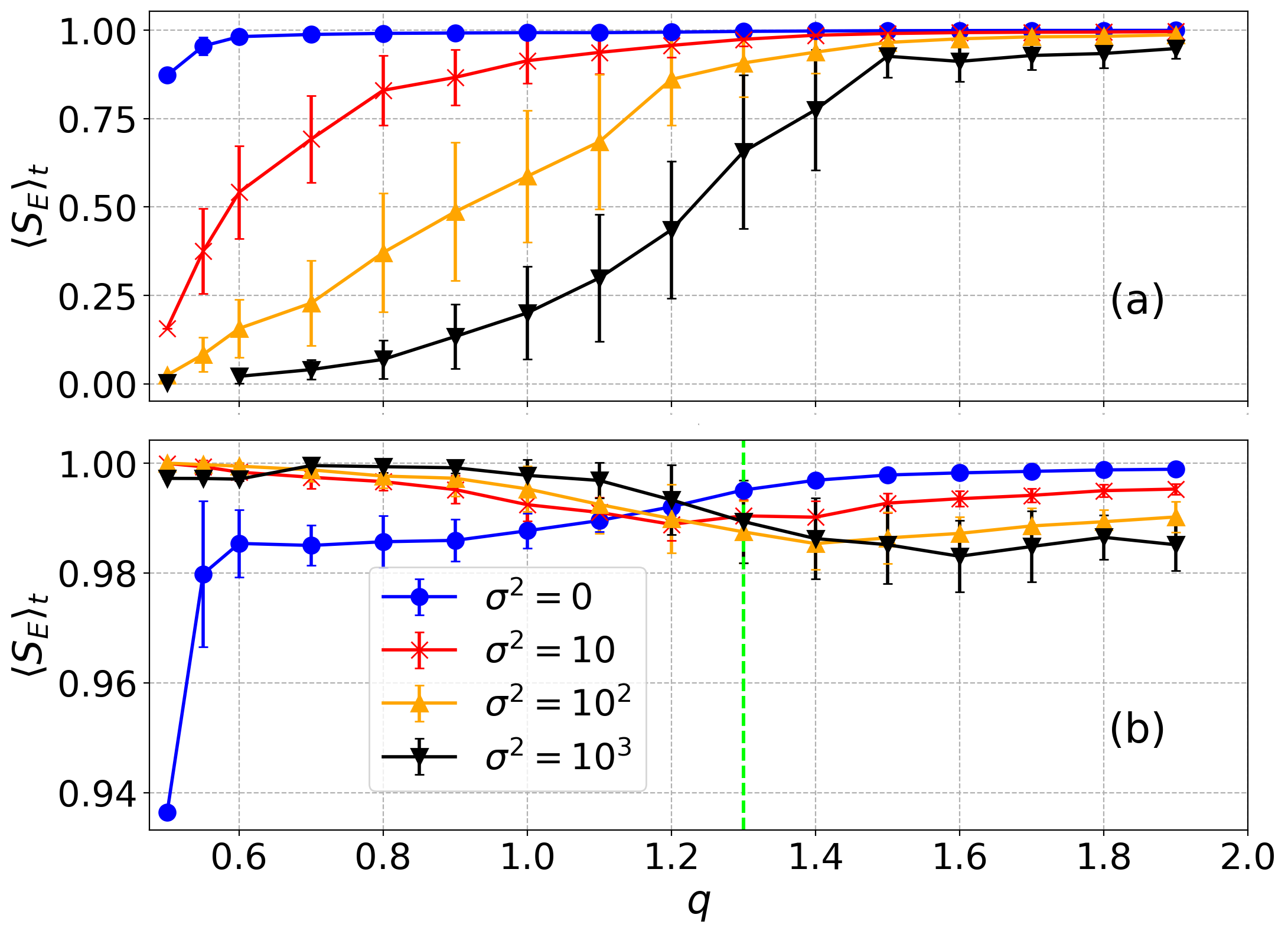}
        \caption{(Color online) Time average coin entanglement entropy as a function of the $q$ parameter in
        the gEQW with initially localized and delocalized position states.
        The coin operator used in (a) was Eq. (\ref{eq:kemp_coin}) with $\theta = \pi/4$ and with 
        $\Omega = \pi/2$ and $\phi = 0$, and in (b) the Hadamard operator with 
        $\Omega = \pi/3$ and $\phi \approx 0.696\pi$. Each data point was obtained through $50$ simulations.}
        \label{fig:me_x_q_gauss}
    \end{figure}

Next, we investigate how the mean entanglement entropy varies as we change $q$ in the $q$-exponential distribution with delocalized initial states.
Fig.~\ref{fig:me_x_q_gauss} depicts the time average entanglement entropy in the generalized elephant quantum walk as a function of $q$ for different position initial variances in (a) with $\Omega = \pi/2$ and $\phi = 0$ and (b) with $\Omega = \pi/3$ and $\phi$ given by Eq. (\ref{eq:orthey_rel}), in the coin initial state. Taking $q = 0.5$, we can see that the time average entanglement indeed decreases as we increase the initial position variance, at least in the case where we use the Kempe coin operator with $\theta = \pi/4$ and $\Omega = \pi/2$ (a).
We note that for $q \in (0.5,1.5]$ the entanglement entropy decreases in comparison with the initially localized case as well; however, there is concomitantly an increase in the uncertainty of the data points. That can be assigned to the fact that the time evolution of the entanglement in the gEQW with $q$ in this region presents very large oscillations, which in our interpretation indicates that with the use of initially delocalized states the walker takes more time to reach the quasi-stationary regime.
Nonetheless, comparing with the deterministic DTQW we have an increase on the coin entropy and by inferring the asymptotic behavior of $\langle S_E \rangle_t$ vs. $q$ we can say that this diminishing goes to zero  as $q \rightarrow \infty$.

Considering the lower panel Fig.~\ref{fig:me_x_q_gauss}(b), we see that the average entropy decreases, but in a smaller degree, as we increase $q$ from $0.5$ in the delocalized cases, being surpassed by the localized ones when $q = 1.3$. As in the Fig.~\ref{fig:me_x_q_gauss}(a) panel, that can be attributed to a delay in the reaching the quasi-stationary regime by the use of delocalized initial states. Bridging those observations with the results obtained in Fig.~\ref{fig:me3d_gauss}(b), it is possible to assert this decrease goes to zero as $q \rightarrow \infty$.

\begin{figure}[!ht]
        \centering
        \includegraphics[scale = 0.315]{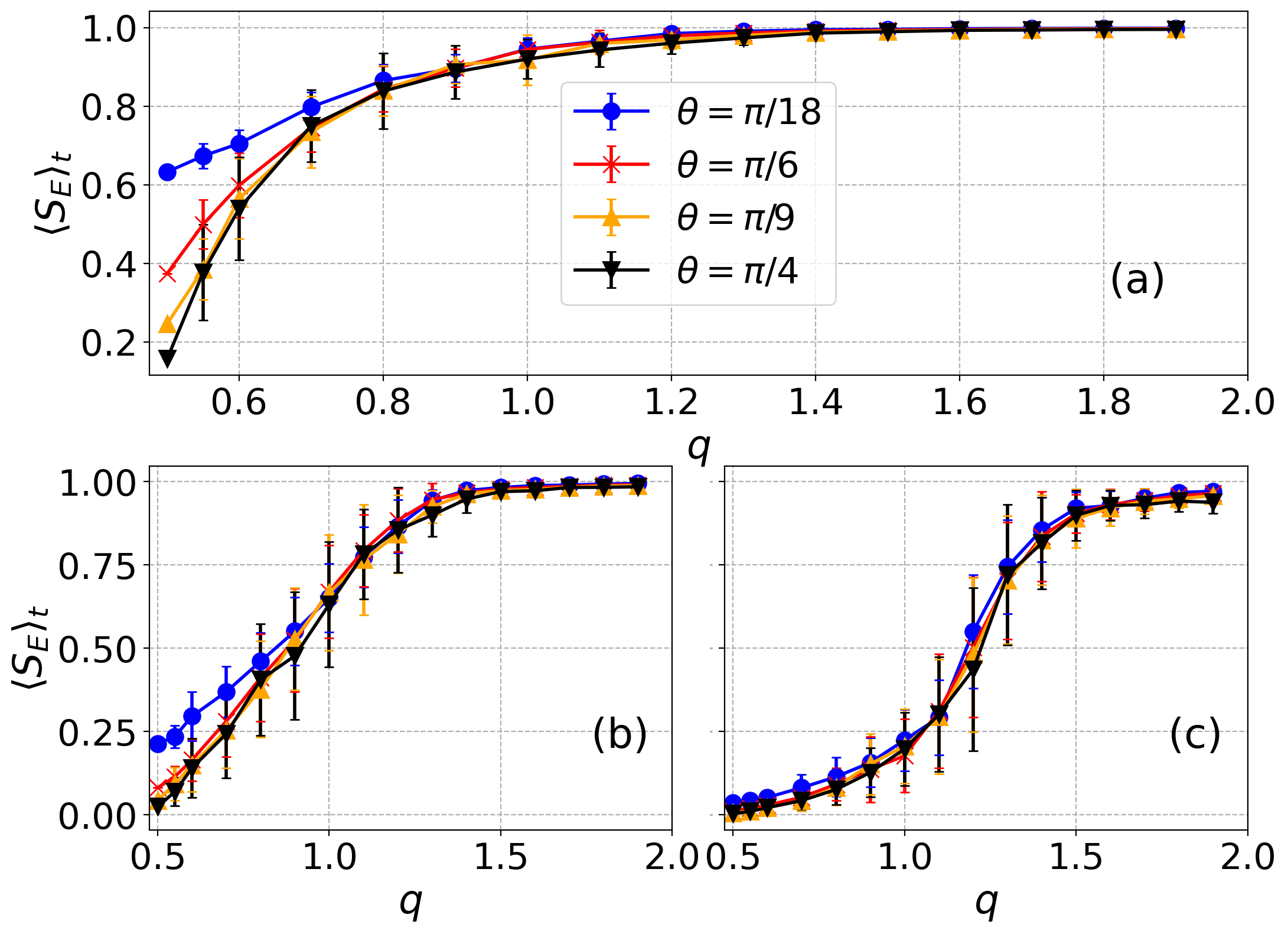}
        \caption{Time average entanglement entropy as a function of $q$ considering different values of $\theta$
        in the Kempe coin $C_k(\theta)$ Eq. (\ref{eq:kemp_coin})
        considering the initially delocalized gEQW, with $\sigma^2 = 10$ (a),
        $\sigma^2= 10^2$ (b) and $\sigma^2 = 10^3$ (c), using the Kempe coin operator. 
        In all simulations a coin initial state was used with $\Omega = \pi/2$  and $\phi = 0$ in Eq. (\ref{eq:coin_instate}).
        Each data point was obtained through $50$ simulations.}
        \label{fig:me_x_q_gauss_theta}
    \end{figure}

Finally, we look at the time-averaged coin entanglement entropy as a function of $q$ for different values of $\theta$ in Eq.(\ref{eq:kemp_coin}) and considering a Gaussian position initial state with $\sigma^2 = 10$ (a), $\sigma^2= 10^2$ (b) and $\sigma^2 = 10^3$ (c) in Fig.~\ref{fig:me_x_q_gauss_theta}. One see that by varying $\theta$ the mean entanglement curve changes significantly only as the initial position variance is low, indicating that with regard to the Kempe coin operator the initial position variance plays a major role in the time average coin entanglement entropy for greater values of $\sigma$.

\begin{figure}[!ht]
        \centering
        \includegraphics[scale = 0.315]{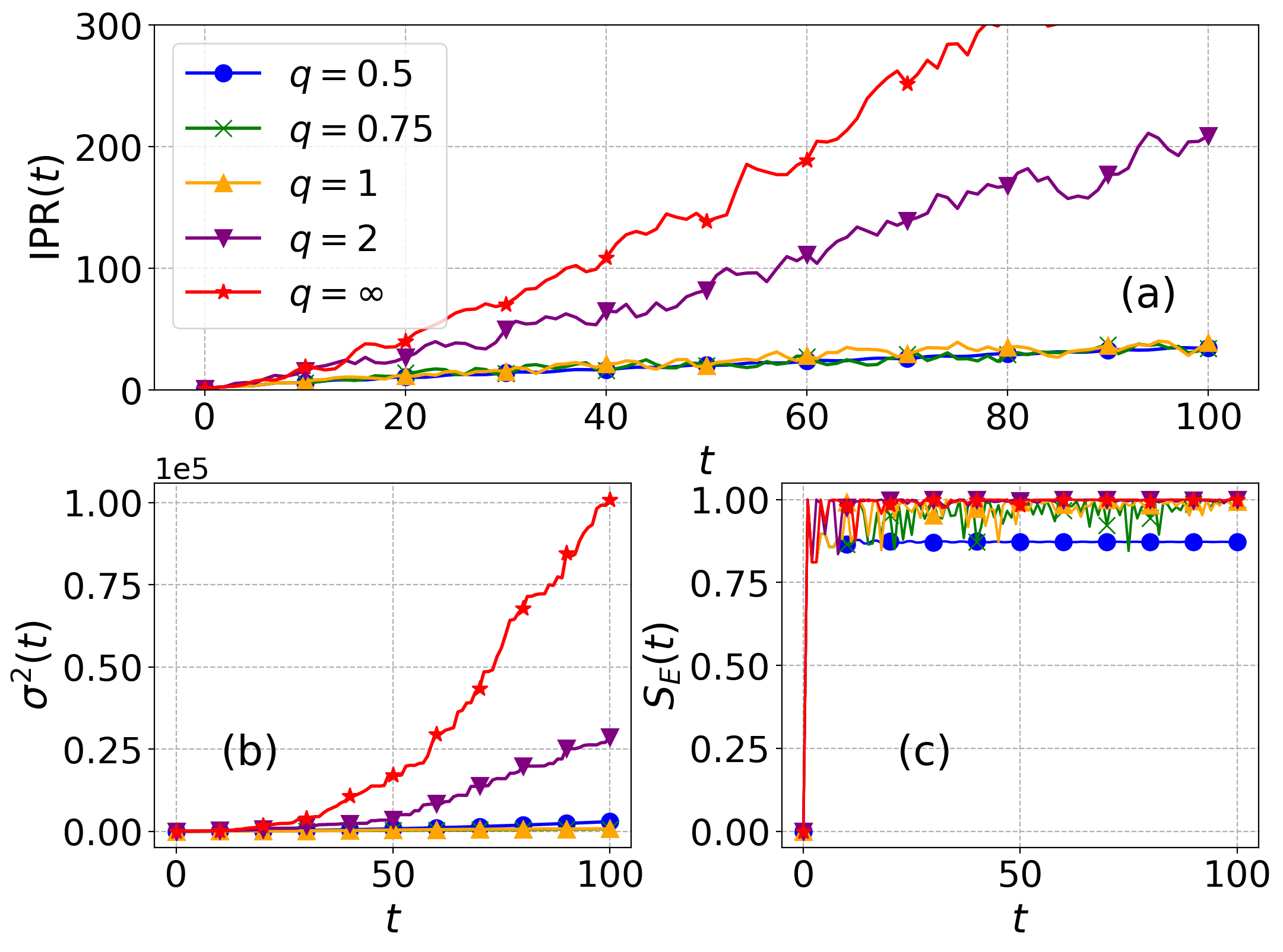}
        \caption{(Color online) IPR time series for different generalized elephant quantum walks $(a)$, variance $(b)$ and von Neumann entropy $(c)$ time evolution for the 
        same gEQWs. The coin operator used was $C_k(\pi/4)$ Eq. (\ref{eq:kemp_coin}) and the initial state considered in all curves was the one localized in the origin with 
        $\phi = 0$ and $\Omega = \pi/2$ in Eq. (\ref{eq:coin_instate}) as coin initial state.}
        \label{fig:IPR}
    \end{figure}

Let us briefly mention that for some important algorithmic applications it is desired to control propagation as well as the the way in which each spinor participates in the 
wavepacket~\cite{kendon2003decoherence,martin2020optimizing}. In this sense we mention that the gEQW is also interesting as can be seen, for instance, in Fig.~\ref{fig:IPR} where we  quantify how much contribution each state provides to the full wavepacket by means of the Inverse Participation Ratio (IPR) of the probability distribution, $IPR \equiv  \left( \sum_x (P_t(x))^2 \right)^{-1}$~\cite{ghosh2014simulating,yalccinkaya2015two,zeng2017discrete,derevyanko2018anderson,buarque2019aperiodic,pires2020quantum}. Such measure has 2 extremes:
(i) fully localized states where $ P_t(x)=\delta_{x,0} $ thus $ IPR=1$;
(ii) fully delocalized states where $  P_t(x)=1/N $  hence $ IPR= N$ where $N$ is the maximum possible number of sites in which $P_t(x)$ can be distributed. 
We see in Fig.~\ref{fig:IPR} that by tuning q it is possible to engineer changes in the  probability distribution $P_t(x)$ in a way that we can control propagation as well as spatial  participation of each spinor in the full wavepacket without reducing the coin-position entanglement.

\subsection{Quasi-stationary regime}

To analyze the long-time behavior of the quantum coin evolution with regard to its state changes we have to use a measure of distinctness between quantum states. For that reason, we have employed the trace distance
    \begin{equation}
        D(\rho,\sigma)= \frac{1}{2}\|\rho - \sigma\|_1
        \mbox{ ,}
    \end{equation}
where $\|A\|_1 = \mbox{tr}\sqrt{AA^{\dagger}}$ is the matrix 1-norm. If $\rho = \sigma$, then the trace distance between them is zero and if they are matrices representing orthogonal states, i.e. $\sigma = \rho^\perp$, their distance is maximum. Hence, by calculating the trace distance between two successive states, $D(\rho_c(t+1),\rho_c(t))$, we can find how, if so, the generalized elephant quantum walk goes to the quasi-stationary regime, that here we define as the dynamics time regime in which the trace distance between two successive states is constant on average, being zero in the limit of a true stationary regime.

\begin{figure}[!ht]
        \centering
        \includegraphics[scale = 0.315]{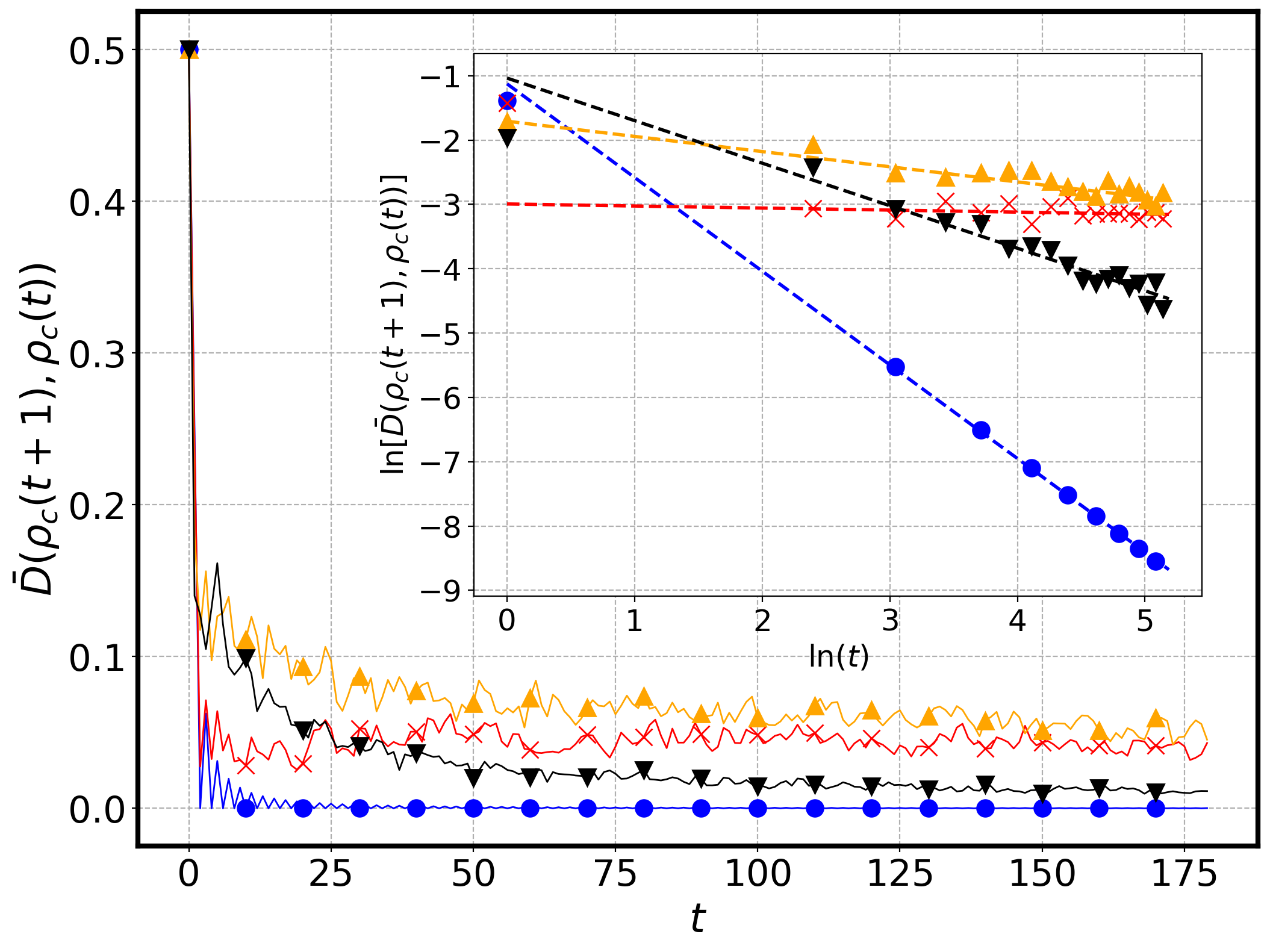}
        \caption{(Color online) Time evolution of the trace distance between two successive coin states for the initially localized gEQW with $q = 0.5$ (blue circle), $q = 0.6$ (red cross), $q = 1$ (orange up triangle) and $q = \infty$ (black down triangle). The coin initial state used was the one following Eq. (\ref{eq:coin_instate}) with $\Omega  = \pi/2$ and $\phi = 0$, using $C_k(\pi/4)$ as coin operator through the evolution. The size of the simulations sample considered for all curves, except $q = 0.5$, was $50$. The inset shows the log-log graph of the same curves, with corresponding decay exponents $ -\beta$,  $(-1.456 \pm 0.004)$ for $q = 0.5$, $(-0.03 \pm 0.02)$ for $q = 0.6$, $(-0.236 \pm 0.008)$ for $q  = 1$ and $(-0.66 \pm 0.01)$ for $ q = \infty$.}
        \label{fig:td_x_q_loc}
    \end{figure}

We begin by looking at the trace distance evolution for different generalized elephant quantum walks, with the standard DTQW included, using initially localized walker states. For $q \ne 0.5$, given that the evolution is stochastic, the trace distance considered is an ensemble average. From Fig.~\ref{fig:td_x_q_loc} we can see that the trace distance decays following a power law in time, $\bar{D} \propto t^{-\beta}$. Also, by increasing the amount of randomness the quantum walk goes to the stationary regime slower than in the deterministic case, with the decay law exponent $\beta$ -- given by the log-log inset fittings -- equal to approximately $1.5$ for the standard DTQW, $\beta \approx 0.03$ for $q = 0.6$, $\beta \approx 0.24$ for $q = 1$ and $\beta \approx 0.66$ for the elephant quantum walk. Moreover, it is possible to affirm that the decay exponent does not follow a simple inverse relationship with the amount of randomness, since the decay exponent for the completely random case is greater than for $q = 0.6$ and $q = 1$.

\begin{figure}[!ht]
        \centering
        \includegraphics[scale = 0.32]{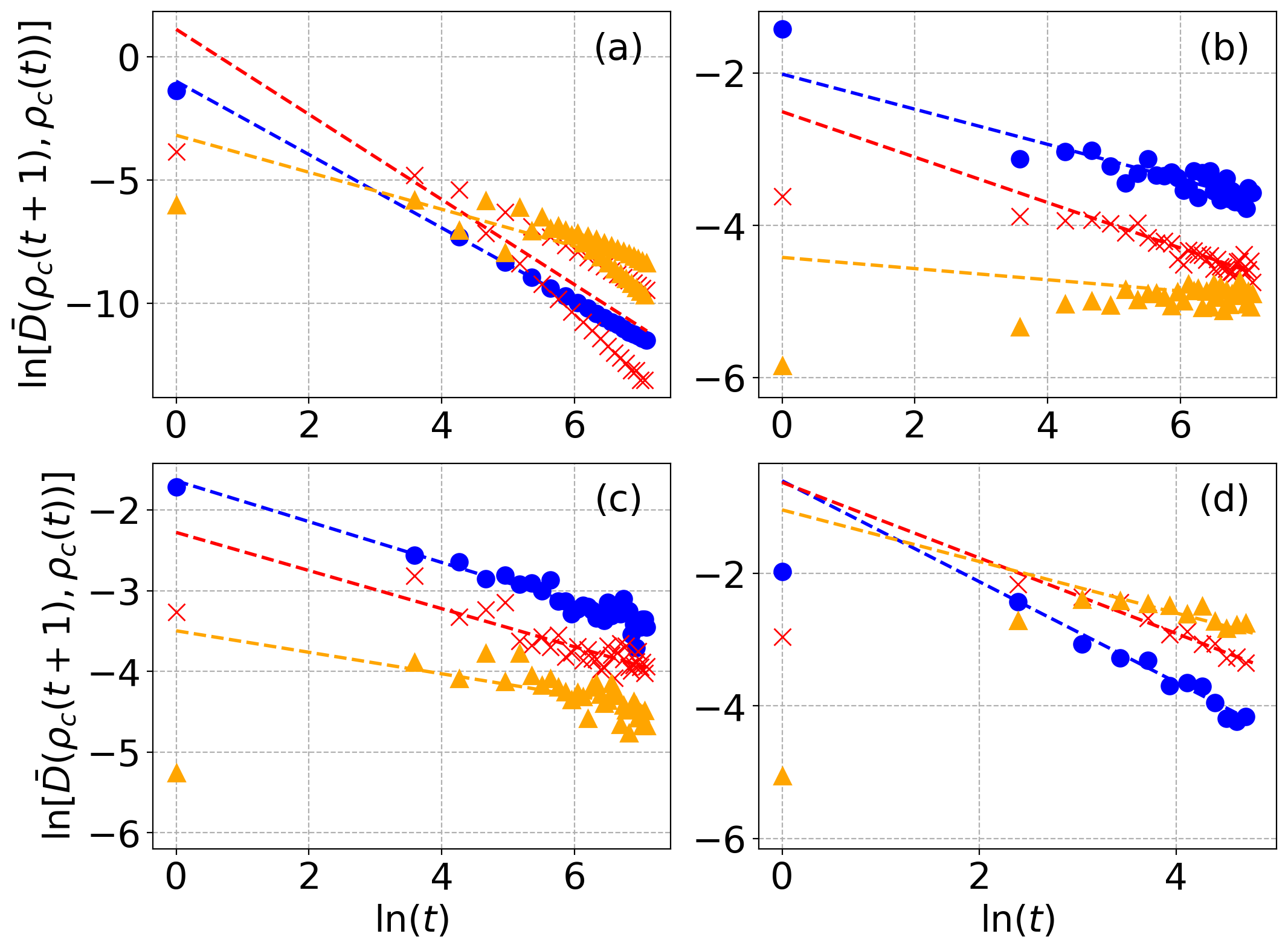}
        \caption{(Color online) Log-log graphs of the average trace distance between two successive coin states time evolution in the generalized elephant quantum walk using $C_k(\pi/4)$ as coin operator with $\phi = 0$ and $\Omega = \pi/2$ determining the coin initial state with different initial variances. 
        The initial variances are $\sigma^2 = 0$ (blue circle), $\sigma^2 = 10$ (red star) and $\sigma^2 = 10^2$ (orange up triangle). 
        In $(a)$ panel we have the standard DTQW, $(b)$ $q = 0.6$, $(c)$ with $q = 1.0$ and $(d)$ corresponding to $q = \infty$. The average was calculated through $50$ simulations for each curve.}
        \label{fig:time_td_comparison}
    \end{figure}

Now we move to see what are the effects of using an initially delocalized state. As a means of comparison, first we look at the standard DTQW trace distance Fig.~\ref{fig:time_td_comparison}(a). It is possible to note that the use of initially delocalized states introduces oscillations and a transient regime in evolution that is made longer when we increase the initial variance. Moreover, by fitting the data points for $t \gg 1$ into a power law and calculating the decay exponents (TAB.~\ref{tab:decay_exps}) we see that by increasing the initial delocalization the quantum walk reaches the quasi-stationary regime faster than in the localized case when $\sigma^2 = 10$ (red cross curve) but slower when $\sigma^2 = 10^2$ (orange up triangle curve). Besides the fact that a true stationary regime does not exists, for the quantum walks with random step sizes, $q = 0.6$ Fig.~\ref{fig:time_td_comparison}(b), $q = 1$ Fig.~\ref{fig:time_td_comparison}(c) and $q = \infty$ Fig.~\ref{fig:time_td_comparison}(d) the same features are observed. With $q = 0.6$, when we use $\sigma^2 = 10$ the quasi-stationary regime is achieved faster than in the localized case, but with $\sigma^2 = 10^2$ it is achieved much more slower, with a longer initial transient increasing. As we increase the amount of randomness this transient takes much more time, as we can note from Fig.~\ref{fig:time_td_comparison}(c-d) and we do not observe a faster decay for $\sigma^2 = 10$ (see TAB.~\ref{tab:decay_exps}).

     \begin{table}[!ht]
        \caption{Table of the decay exponent $\beta$ of the trace distance between two time successive states considering the different values of $q$
        and initial variance $\sigma^2$ obtained through the fittings of the curves for $t \gg 1$ in Fig.~\ref{fig:time_td_comparison}.}
        \label{tab:decay_exps} 
        \begin{ruledtabular}
        \centering
        \begin{tabular}{cccc}
            \backslashbox{$q$}{$\sigma^2$} & $0$ & $10$ & $10^2$ \\
            \colrule
            $0.5$ & $1.487 \pm 0.001$ & $1.73 \pm 0.04$ & $0.75 \pm 0.02$\\
            $0.6$ & $0.23 \pm 0.01$  & $0.30 \pm 0.01$ & $0.07 \pm 0.01$ \\
            $1$ & $0.253 \pm 0.003$ & $0.236\pm 0.005$ & $0.133 \pm 0.006$ \\
            $\infty$ & $0.76 \pm 0.02$ & $0.57 \pm 0.02$ & $0.37 \pm 0.03$ \\
        \end{tabular}
    \end{ruledtabular}
     \end{table}
    
This property of retarding the quasi-stationary regime as one increases the initial delocalization explains the greater uncertainty and lower values of the average entanglement of Fig.~\ref{fig:me_x_q_gauss} when one also increases the amount of randomness. It is also remarkable that the feature of extending the transient regime was previously observed in quantum walks with dynamically random coin operators \cite{vieira2013dynamically}, where the decaying trace distance follows a power law with exponent equal to $-1/4$. This tell us that this property is indeed a feature of dynamically random quantum walks, now including the use of random shift operators. We highlight that the figures \ref{fig:td_x_q_loc}-\ref{fig:time_td_comparison} can also be used as evidence that in average the quantum walk with random steps sizes indeed have a quasi-stationary regime, with some of them taking more time than others to reach it depending on the initial delocalization and degree of randomness of the step sizes.

\section{Final remarks}
\label{sec:remarks}

We have analyzed the production of coin entanglement entropy in the generalized elephant quantum walk considering different types of initial conditions and coin operators. 

First, looking at the initially localized walker state, we have observed that the time average coin entanglement reaches its maximum value for almost all parameters in the Kempe coin operator and polar angles in the coin initial state Bloch sphere when the step distribution is uniform and exponential [Fig.~\ref{fig:mentang}(a) and (c)], i.e. with $q = \infty$ and $q = 1$ in the distribution parameter respectively. That behavior does not occur in the standard discrete time quantum walk considering the same parameters, as presented in Fig.~\ref{fig:mentang}(b). On the other hand, when we have used a more general type of coin operator for the same initial walker state, Eq.~(\ref{eq:u2_coin}), 
we learned the elephant quantum walk entanglement does not change [Fig.~\ref{fig:mentang_3d_beta}(b)], while in the DTQW 
[Fig.~\ref{fig:mentang_3d_beta}(a)] the time average entanglement entropy can vary from its value given when we use $\theta = \pi/4$ in the Kempe coin up to almost the maximum value. Looking at the $\langle S_E \rangle_t$ vs. $q$, Figs.~\ref{fig:me_x_q} and \ref{fig:me_x_q_beta}, we have seen that it only takes a small amount of disorder in order to greatly improve the time average entanglement, going from $\langle S_E \rangle_t \approx 0.8724$ with only unit step sizes to $0.9852$ with probability of approximately $6\%$ of having steps
of sizes equal to $2$ ($q = 0.6$), for $\theta = \pi/4$. By changing the Kempe coin parameter, one only increases the initial entanglement average as a function of $q$ increase rate and the same goes for the Bloch polar angle of the coin initial state.

Next, we have considered the use of Gaussian delocalized initial states, and the use of random steps sizes also improved the time average
coin entanglement when one uses the Kempe coin operator and $\Omega = \pi/2$ and $\phi = 0$ in the coin initial state.
However, by analyzing the dependence of the average entanglement on the amount of disorder we have learnt that for $q \in (0.5,1.5]$ the uncertainty on the data increased, indicating a possible retarding on reaching the quasi-stationary regime. This have been confirmed by an analysis of the quasi-stationary regime through the trace-distance between two time successive coin states. By comparing the initially localized and delocalized walks, we found that when one increases the initial randomness the initial transient regime becomes longer. Looking at the stationary regime, the power law exponent describing the trace distance decay, in both localized and delocalized walks, becomes greater than the deterministic case, something that was also observed in quantum walks with dynamically random coin operators.
Nonetheless, we assert that for almost all initial coin states and coin operators, the generalized elephant quantum walk enhances the coin entanglement entropy for delocalized initial states taking it to the supreme as $q \rightarrow \infty$.

Although it is usually expected that disorder weakens quantum features, nowadays it has already been established by numerical~\cite{chandrashekar2012disorder}, theoretical~\cite{vieira2013dynamically} and experimental~\cite{wang2018dynamic,tao2021experimental} work that dynamical disorder embedded in the coin operator of QWs acts as a maximal entanglement generator.  Such strengthening of the entanglement takes place at the cost of  weakening the controllability of the spreading features. In sharp contrast, novel properties emerge when the dynamical disorder is embedded in the shift operator of QWs \cite{di2018elephant}. The generalized elephant quantum walk  (gEQW)\cite{pires2019multiple}  is a protocol that has a remarkable variety of 
scaling behavior (diffusive, superdiffusive, ballistic, hyperballistic, as shown in Fig.\ref{fig:mean_alpha_x_q})  and still produces maximally entangled coin states. 
Such features make the gEQW an interesting protocol for potential applications demanding controllability of transport properties whilst keeping a maximum asymptotic entanglement. However, in order to move on towards future applications, it is necessary to understand whether the enhancement of the entanglement of the gEQW is robust or only valid for the specific coin operators and local states studied in Ref.~\cite{pires2019multiple}. 
Here, we have shown that such highly entangled states are also achieved with delocalized initial states and general coin operators in a robust way.


That been said, we conclude that
disordered quantum walks -- either with dynamical disorder in the coin operator or in the shift operator -- generate maximally entangled coin states for almost all initial coin states and coin operators considering initially localized walkers and for the delocalized ones in the limit $q \rightarrow \infty$ the same is guaranteed.

In summary, we have extensively shown that the application  of dynamically disordered QWs as a maximal entanglement generator is also possible when: (i) the temporal disorder is embedded in the step operator and (ii) general coin operators and  both local and delocalized initial states are used.

We highlight that differently from all previous QW-based protocols for generating highly entangled states -- that do not allow high controllability of the spreading -- our protocol opens the doors for potential new possibilities of applications of dynamically disordered QWs as a robust maximal entanglement generator in programmable setups that ranges from slower-than-ballistic to faster-than ballistic.

In quantum simulations, the primary objective is to engineer a quantum system that could be tuned to model the properties of other quantum systems.  In this sense, it is evident from our work that the rich spreading phenomenology of the gEQW accompanied by a robust amplification entanglement captures the essence of a programmable quantum system.

 In future works, we aim at investigating to what extent it is possible to optimize the efficiency of our setup for generating highly entangled states without losing the richness of the spreading behavior (diffusive, superdiffusive, ballistic, hyperballistic).

\begin{acknowledgments}
CBN acknowledges the support from IFSC-USP and from Coordena\c{c}\~{a}o de Aperfei\c{c}oamento de Pessoal de N\'{i}vel Superior - Brasil (CAPES) - Finance Code 001. MAP acknowledges the support by the funding agency FUNCAP. DOSP acknowledges the support by the Brazilian funding agencies CNPq (Grant No. 307028/2019-4), FAPESP (Grant No. 2017/03727-0), and the Brazilian National Institute of Science and Technology of Quantum Information (INCT/IQ). SMDQ acknowledges the financial support from CNPq Grant No. 307028/2019-4).
\end{acknowledgments}


\providecommand{\noopsort}[1]{}\providecommand{\singleletter}[1]{#1}

\end{document}